\documentclass[a4paper,11pt]{article}
\pdfoutput=1 % if your are submitting a pdflatex (i.e. if you have
\usepackage{jheppub} % for details on the use of the package, please
\usepackage[T1]{fontenc} % if needed
\usepackage{graphicx,xcolor}
\usepackage{multicol}
\usepackage{csquotes}
\usepackage{ulem}

\title{\boldmath Causality and stability in relativistic viscous non-resistive magneto-fluid dynamics}

\author[a]{Rajesh Biswas\note{Corresponding author.}}
\author[a]{, Ashutosh Dash}
\author[a]{, Najmul Haque}
\author[b]{, Shi Pu}
\author[a,1]{, Victor Roy}
\affiliation[a]{School of Physical Sciences, National Institute of Science Education and Research, HBNI, 752050,\\ Jatni , India.}
\affiliation[b]{Department of Modern Physics, University of Science and Technology of China, Hefei 230026, China.}

\emailAdd{rajeshphysics143@gmail.com}
\emailAdd{ashutosh.dash@niser.ac.in}
\emailAdd{nhaque@niser.ac.in}
\emailAdd{shipu@ustc.edu.cn}
\emailAdd{victor@niser.ac.in}

\abstract{
We investigate the causality and the stability of the relativistic viscous magneto-hydrodynamics in the framework of the Israel-Stewart (IS) second-order theory, and also within a modified IS theory which incorporates the effect of magnetic fields in the
relaxation equations of the viscous stress. We compute the dispersion relation by
perturbing the fluid variables around their equilibrium values. In the ideal magnetohydrodynamics limit, the linear dispersion relation yields the well-known propagating modes: the Alfv\'en and the magneto-sonic modes. 
In the presence of bulk viscous pressure,  the causality bound is found to be independent of the magnitude of the magnetic field. The same bound also remains true, when we take the full non-linear form of the equation using the method of characteristics. In the presence of shear viscous pressure,  the causality bound is independent of the magnitude of the magnetic field for the two magneto-sonic modes. The causality bound for the shear-Alfv\'en modes, however, depends both on the magnitude and the direction of the propagation. For modified IS theory in the presence of shear viscosity, new non-hydrodynamic modes emerge but the asymptotic causality condition is the same as that of IS. In summary, although the magnetic field does influence the wave propagation in the fluid, the study of the stability and asymptotic causality conditions in the fluid rest frame shows that the fluid remains stable and causal given that they obey certain asymptotic causality condition.}

\begin{document} 
\maketitle
\flushbottom
\section{Introduction}
\label{sec:intro}

%%\begin{figure}[tbp]
%%\centering % \begin{center}/\end{center} takes some additional vertical space
%%\includegraphics[width=.45\textwidth,trim=0 380 0 200,clip]{img1.pdf}
%%\hfill
%%\includegraphics[width=.45\textwidth,origin=c,angle=180]{img2.pdf}
% "\includegraphics" is very powerful; the graphicx package is already loaded
%%\caption{\label{fig:i} Always give a caption.}
%%\end{figure}

%%\begin{table}[tbp]
%%\centering
%%\begin{tabular}{|lr|c|}
%%\hline
%%x&y&x and y\\
%%\hline 
%%a & b & a and b\\
%%1 & 2 & 1 and 2\\
%%$\alpha$ & $\beta$ & $\alpha$ and $\beta$\\
%%\hline
%%\end{tabular}
%%\caption{\label{tab:i} We prefer to have borders around the tables.}
%%\end{table}

%%%%%%% huge magnetic feilds are generated in heavy ion collisions

In relativistic heavy-ion collisions experiment, two fast moving charged nuclei collide with each other and generate a deconfined state of matter known as Quark-Gluon-Plasma~(QGP). In non-central collisions an extremely strong magnetic field~($\sim 10^{18}\mbox{-}10^{19}$ Gauss) is also produced in the initial stages refs.~\cite{Bzdak:2011yy, Deng:2012pc,Tuchin:2013ie,Roy:2015coa,Li:2016tel} mostly due to the spectator protons.

%%%%%%%CME, CSE and other important quantum transport phenomena

The huge magnetic fields induce many novel quantum transport phenomena. One of the most interesting and important phenomena is the Chiral Magnetic Effect~(CME) refs.~\cite{Kharzeev:2004ey,Kharzeev:2007jp,Fukushima:2008xe}, which means a charge current will be induced and be parallel to the magnetic fields in a chiraly imbalanced system. 
Along with the CME, it was also theoretically predicted that massless fermions with the same charge but different chirality will be separated, known as chiral separation effect (CSE). The electric fields may also cause the chiral separation effects or chiral Hall effects refs.~\cite{Huang:2013iia,Pu:2014cwa, Pu:2014fva, Jiang:2014ura}. There are many discussions on other high order non-linear chiral transport phenomena refs.~\cite{ Satow:2014lia, Chen:2016xtg, Ebihara:2017suq}. 
One theoretical framework for studying these quantum transport phenomena is the chiral kinetic theory refs.~\citep{Stephanov:2012ki,Son:2012zy,Chen:2012ca,Manuel:2013zaa,Manuel:2014dza,Chen:2014cla,Chen:2015gta,Hidaka:2016yjf,Mueller:2017lzw,Hidaka:2017auj,Hidaka:2018mel,
Huang:2018wdl,Gao:2018wmr,Liu:2018xip,  Lin:2019ytz, Lin:2019fqo} and numerical simualations based on this framework can be found in refs.~\citep{Sun:2016nig,Sun:2016mvh,Sun:2017xhx,Sun:2018idn,Sun:2018bjl,Zhou:2018rkh,Zhou:2019jag,Liu:2019krs}. Recently, the chiral particle production is found to be connected to the famous Schwinger mechanism ref.~\cite{Fukushima:2010vw}, and is proved through the world-line formalism ref.~\cite{Copinger:2018ftr} and Wigner functions ref.~\cite{Sheng:2018jwf}. There are also many theoretical studies of CME from the quantum field theory refs.~\cite{Feng:2017dom, Wu:2016dam, Lin:2018aon, Horvath:2019dvl, Feng:2018tpb} and the chiral charge fluctuation refs.~\cite{Hou:2017szz, Lin:2018nxj}. The strong magnetic field might also induces anisotropic transport of momentum which results in the anisotropic transport coefficients refs.~\cite{Dash:2020vxk,Kurian:2020kct}. In refs.~\cite{Voronyuk:2011jd,  Greif:2014oia} relativistic Boltzmann equation was used to study the effect of electromagnetic fields in heavy-ion collisions. For the recent developments, one can see the reviews refs.~\citep{Kharzeev:2015znc,Liao:2014ava,Miransky:2015ava,Huang:2015oca,Fukushima:2018grm,Bzdak:2019pkr,Zhao:2019hta, Liu:2020ymh,Gao:2020vbh} and references therein.

%%%%%%%%development of MHD and anaylatic solutions
The charge separation in Au+Au collisions are claimed to be observed by the STAR collaboration refs.~\citep{Abelev:2009ac,Abelev:2009ad,Abelev:2012pa}. However, it is still a challenge to extract the CME signals from the huge backgrounds caused by the collective flows refs.~\citep{Khachatryan:2016got,Sirunyan:2017quh,Sirunyan:2017tax}. Therefore, it requires the systematic and quantitative studies of the evolution of the QGP coupled with the electromagnetic fields for the discovery of CME. 
It is widely accepted that the QGP produced in high energy heavy-ion collisions behaves as almost ideal fluid (i.e., possess very small shear and bulk viscosity). This conclusion was made primarily based on the success of relativistic viscous 
   hydrodynamics simulations in explaining a multitude of experimental data with a very small specific shear 
   viscosity ($\eta$/s) as an input refs.~\cite{Shen:2011zc,Luzum:2008cw,Heinz:2013th,Bozek:2012qs,Roy:2012jb,Heinz:2011kt,Niemi:2012ry,Schenke:2011bn}. Most of these theoretical studies use IS second-order causal viscous hydrodynamics 
 formalism or some variant of it.  The fact that the QGP is composed of electrically charged quarks indicates that 
 it should have finite electrical conductivity which is corroborated by the lattice-QCD calculations refs.~\cite{Gupta:2003zh,Aarts:2014nba,Amato:2013naa}  and perturbative QCD calculations refs.~\cite{Arnold:2003zc,Chen:2013tra}. 
The electrical conductivity of the QGP and the hadronic phase was also calculated by various  other groups (mostly using the Boltzmann transport equation) see refs.~\citep{Dey:2019vkn,Dey:2019axu,Das:2019pqd,Harutyunyan:2016rxm,Kerbikov:2014ofa, Nam:2012sg,Huang:2011dc,Hattori:2016lqx,Kurian:2018qwb,Kurian:2017yxj,Feng:2017tsh,Fukushima:2017lvb,Das:2019wjg,Das:2019ppb,Ghosh:2019ubc}. It is then natural to expect that the appropriate equation of motion of the high temperature QGP and  low temperature  hadronic phase under large magnetic fields is given by the relativistic viscous magneto-hydrodynamic framework. As mentioned earlier the IS second-order theory of causal dissipative fluid dynamics, although successful, known to allow superluminal signal propagation (and hence acausal) under  certain circumstances refs.~\cite{Hiscock:1985zz,Pu:2009fj,Denicol:2008ha,Floerchinger:2017cii}. It is then important to know under what physical conditions the theory remains causal and stable in presence of a magnetic field which is also important for the numerical magnetohydrodynamics~(MHD) studies of heavy-ion collisions.\par
Relativistic magnetohydrodynamics~(RMHD) is a self-consistent macroscopic framework which describe the evolution of any charged fluid in the presence of electromagnetic fields refs.~\cite{Roy:2015kma,Pu:2016ayh,Hongo:2013cqa,Inghirami:2016iru,Inghirami:2019mkc,Siddique:2019gqh,Wang:2020qpx}. In ref.~\cite{Roy:2015coa}, we have computed the ratio of the magnetic field energy to the fluid energy density in the transverse plane of Au+Au collisions at $\sqrt{s_{NN}}  = 200~\textrm{GeV}$ in the event-by-event simulations. Our results imply that the magnetic field energy is not negligible. In ref.~\cite{Roy:2015kma}, we have derived the analytic solutions of a longitudinal Bjorken boost invariant MHD with transverse electromagnetic fields in the ideal limit. We have found that the transverse magnetic fields will decay as $\sim 1/\tau$ with $\tau$ being the proper time. Later, in ref.~\cite{Pu:2016ayh}, we have studied the corrections from the magnetization effects and extended the discussion to~($2+1$)-dimensional ideal MHD refs.~\cite{Pu:2016bxy,Pu:2016rdq}. We have also investigated the effects of large magnetic fields on $(2+1)$-dimensional reduced MHD at $\sqrt{s_{NN}} = 200~\textrm{GeV}$ ref.~\cite{Roy:2017yvg}. Very recently, we have derived the analytic solutions of  MHD in the presence of finite electric conductivity, CME and chiral anomaly ref.~\cite{Siddique:2019gqh} and extended the results to cases with the transverse and longitudinal electric conductivities ref.~\cite{Wang:2020qpx}. For numerical simulations of ideal MHD, one can see refs.~\cite{Inghirami:2016iru,Inghirami:2019mkc}.

%%%%%%%%%%%Causality and stability, the motivation
As mentioned earlier in the ordinary relativistic hydrodynamics, the widely used framework is the second order IS theory~\cite{Israel:1979wp}. The pioneering studies on the instabilities of first order hydrodynamics are shown in refs.~\cite{Hiscock:1983zz,Hiscock:1985zz}. Later, the systematic studies for the dissipative fluid dynamics have been done earlier with bulk viscous pressure~\cite{Denicol:2008ha}, shear viscous stress~\cite{Pu:2009fj} and heat currents~\cite{Pu:2010zz}, also see refs.~\cite{Denicol:2008rj,Floerchinger:2017cii,Bemfica:2019cop,Bemfica:2020xym}. There have been several recent studies on casualty and stability of ideal MHD in refs.~\cite{Dionysopoulou:2012zv,Grozdanov:2016tdf,Hernandez:2017mch} and reference therein. The extension of MHD to the IS formalism through the help of AdS/CFT has been recently been done in refs.~\cite{Grozdanov:2018fic,Grozdanov:2017kyl}.
%it is still of importance to know under what physical conditions the dissipative theory remains causal and stable in presence of a magnetic field while using the same in the numerical studies for heavy-ion collisions.

We aim to study the stability and causality of the IS theory for MHD, whose form is derived by the complete moment expansion as done in refs.~\cite{Denicol:2018rbw,Denicol:2019iyh}. 
First, we analyze the propagating modes in ideal non-resistive MHD. Next, we discuss the causality and stability of the relativistic MHD with dissipative effects. To analyse the causality and stability of the relativistic viscous fluid, we linearise the relevant equations by using a small sinusoidal perturbation around the local equilibrium and study the corresponding dispersion relations in line with the studies in refs.~\cite{Hiscock:1983zz, Hiscock:1985zz,Denicol:2008ha,Pu:2009fj}. 

%%%%%%%%Structurer of the work
The manuscript is organized as follows: In section~\ref{sec2:formalism} we briefly discuss the energy-momentum tensor of fluid for ideal MHD case and the modified IS theory. In section~\ref{sec3:drel_withoutB} we revisit the standard analysis of causality and stability of a system without magnetic fields. Then, in section~\ref{sec4:ResultDis} we show the stability and causality of an ideal MHD and carry out the analysis of characteristic velocities in section~\ref{sec5:Charvel}. In section~\ref{sec6:NRMHD} we consider the newly developed IS theory for non-resistive MHD. Finally, we conclude our work in section~\ref{sec:conclusion}.

Throughout the paper, we use the natural unit and the flat space-time metric \( g^{\mu\nu}=\text{diag} ({+1,-1,-1,-1})\). The fluid velocity satisfies $u^\mu u_\mu=1$ and the projection operator perpendicular to $u^{\mu}$ is $\Delta^{\mu \nu}\equiv g^{\mu \nu}-u^\mu u^\nu$. The operators $D$ and $\nabla^\mu$ are defined as $D\equiv u^{\mu} \partial_{\mu}$ and $\nabla^{\mu}\equiv \Delta^{\mu \nu} \partial_{\nu}$, respectively.

%%%%%%%%%%%%%%%%%%%%%%%%%%%%%%%%%%%%%%%%%%%%%%%%%%%%%%%%%%%%%%%%%%%%%%%%

\section{Causal relativistic fluid in presence of magnetic field}
\label{sec2:formalism}

In this work we consider the causal relativistic second order theory for relativistic fluids by Israel-Stewart (IS)  and also a modified form of the IS theory in presence of a magnetic field given in ref.~\cite{Denicol:2018rbw}, for later use we define it as NRMHD-IS theory~(here NRMHD corresponds to non resistive magneto-hydrodynamics).
%We will briefly review a relativistic viscous fluid in presence of  electromagnetic fields in the non-resistance limit. 
The total energy-momentum tensor of the fluid can be written as 
\begin{equation}
T^{\mu \nu}=\left(\varepsilon+P+\Pi+B^2\right)u^\mu u^\nu-\left(P+\Pi+\frac{B^2}{2}\right)g^{\mu\nu}-B^{\mu}B^{\nu}+\pi^{\mu\nu},
\label{eq:EMTensFull}
\end{equation}
where $\varepsilon$, $P $ are fluid energy density, pressure , $ u^{\mu}$ is the fluid four velocity and
$\Pi,\pi^{\mu\nu}$ are bulk viscous pressure and shear viscous tensor, respectively. The magnetic and electric four vectors are defined as 
\begin{equation}
B^{\mu}= \frac{1}{2}\epsilon^{\mu\nu\alpha\beta}u_{\nu}F_{\alpha\beta},  \; \; E^\mu=F^{\mu\nu} u_\nu,
\end{equation}
where $F^{\mu\nu}=(\partial^{\mu}A^{\nu}-\partial^{\nu}A^{\mu})$ is the field strength tensor. 
The space-time evolution of the fluid and magnetic fields are described by the energy-momentum conservation 
\begin{equation}
\partial_{\mu}T^{\mu\nu}=0, \label{eq:energy_conservation}
\end{equation}
coupled with Maxwell's equations
\begin{eqnarray}
\partial_\mu F^{\mu\nu}&=& j^{\nu}, \nonumber \\
\epsilon^{\mu\nu\alpha\beta}\partial_{\beta}F_{\nu\alpha}&=&0.
\end{eqnarray}
The non-resistance limit means the electric conductivity $\sigma_e$ is infinite. In this limit, in order to keep the charge current $j^\mu = \sigma_e E^\mu$ be finite, the $E^\mu\rightarrow 0$. Then, the relevant Maxwell's equations which govern the evolution of magnetic fields in the fluid is 
\begin{equation}
\partial_{\nu}(B^{\mu}u^{\nu}-B^{\nu}u^{\mu})=0. 
\end{equation} 
For simplicity,  we will also neglect the magnetisation of the QGP,  which implies an isotropic pressure and no change in the Equation of Sate (EoS) of the fluid due to magnetic field~(e.g. see ref.~\cite{Pu:2016ayh}).\par
In the original IS theory the viscous stresses $\Pi,\pi^{\mu\nu}$ are considered  as an independent dynamical variables given by the following equations~(e.g. see refs.~\cite{Baier:2007ix, Betz:2008me, Betz:2009zz})
\begin{eqnarray}
\Pi &=&\Pi_{\mathrm{NS}}-\tau_{\Pi} \dot{\Pi} \nonumber\\
& &+\tau_{\Pi q} q \cdot \dot{u}-\ell_{\Pi q} \partial \cdot q-\zeta \hat{\delta}_{0} \Pi \theta \nonumber\\
& &+\lambda_{\Pi q} q \cdot \nabla \alpha+\lambda_{\Pi \pi} \pi^{\mu \nu} \sigma_{\mu \nu}, \\ 
\pi^{\mu \nu} &=&\pi_{\mathrm{NS}}^{\mu \nu}-\tau_{\pi} \dot{\pi}^{<\mu \nu >} \nonumber\\
& & +2 \tau_{\pi q} q^{<\mu} \dot{u}^{\nu>}+2 \ell_{\pi q} \nabla^{<\mu} q^{\nu>}+2 \tau_{\pi} \pi_{\lambda}^{<\mu} \omega^{\nu>\lambda}-2 \eta \hat{\delta}_{2} \pi^{\mu \nu} \theta \nonumber\\
& &-2 \tau_{\pi} \pi_{\lambda}^{<\mu} \sigma^{\nu>\lambda}-2 \lambda_{\pi q} q^{<\mu} \nabla^{\nu>} \alpha+2 \lambda_{\pi \Pi} \Pi \sigma^{\mu \nu},
\label{eq:ISFull}
\end{eqnarray}
where $\zeta$ and $\eta$ are bulk and shear viscosity, respectively. The coefficients  $\tau_{\Pi}$ and $\tau_{\pi}$ are the  relaxation times for the bulk and shear viscosity, respectively and $\omega^{\mu \nu} \equiv \frac{1}{2} \Delta^{\mu \alpha} \Delta^{\nu \beta}(\partial_{\alpha} u_{\beta}-\partial_{\beta} u_{\alpha})$ is the vorticity tensor. The subscript NS means the Navier-Stokes values and can be written as 
\begin{eqnarray}
\Pi_{NS} & = & -\zeta \theta = -\zeta\partial_{\mu} u^{\mu}, \nonumber \\
\pi^{\mu\nu}_{NS} & = &2\eta\sigma^{\mu\nu},
\end{eqnarray}
where 
\begin{equation}
\sigma^{\mu \nu}=\nabla^{<\mu} u^{\nu>}=\frac{1}{2}\left(\nabla^{\mu} u^{\nu}+\nabla^{\nu} u^{\mu}\right)-\frac{1}{3} \Delta^{\mu \nu} \partial_{\alpha} u^{\alpha}.
\end{equation}
Note that all of these coefficients 
%$\tau_{\Pi q}, \ell_{\Pi q}, \hat{\delta}_{0}, \lambda_{\Pi q}, \lambda_{\Pi \pi}, \tau_{\pi q}, \ell_{\pi q}, \hat{\delta}_{2}, \lambda_{\pi q},\lambda_{\pi \Pi}$ 
are functions of baryon chemical potential~($\mu$) and temperature~($T$). Equation~\eqref{eq:ISFull} can be derived from the kinetic theory via complete moment expansion, one can see refs.~\cite{Denicol:2012cn, Denicol:2012es, Molnar:2013lta} for more details. 

For further simplification, we also ignore the coupling of viscosity with other dissipative forces and concentrate on  
the following terms
\begin{eqnarray}\label{eq:ISshortbulk}
\Pi &=&\Pi_{\mathrm{NS}}-\tau_{\Pi} \dot{\Pi},\\
\pi^{\mu \nu} &=&\pi_{\mathrm{NS}}^{\mu \nu}-\tau_{\pi} \dot{\pi}^{<\mu \nu>}.\label{eq:ISshortShear}
\end{eqnarray} 
We note that in principle the magnetic field may cause viscous tensor to be anisotropic as shown in ref.~\cite{Huang:2009ue} but in this work we consider zero magnetisation and hence use eqs.~\eqref{eq:ISshortbulk},~\eqref{eq:ISshortShear} for simplicity.

%%%%%%%%%%%%%%%%%%%%%%%%%%%%%%%%%%%%%%%%%%
\section{Dispersion relation in the absence of magnetic field}
\label{sec3:drel_withoutB}

As is known, IS theory is a consistent fluid dynamical prescription which preserves causality provides that the relaxation time associated with the dissipative quantities (such as shear and bulk viscous stresses) are not too small refs.~\cite{Hiscock:1983zz, Hiscock:1985zz,Denicol:2008ha,Pu:2009fj,Pu:2010zz,Denicol:2008rj, Floerchinger:2017cii}. Here we aim to study the stability and causality of a relativistic viscous fluid (governed by the IS equations) in an external magnetic field by linearising the governing equations under a small perturbation. 

Before discussing the causality and stability of a relativistic viscous fluid in a magnetic field, for the sake of completeness, let us summaries here the findings without the magnetic field. We note that the following results are not new and most of them can be found in refs.~\cite{Pu:2009fj, Denicol:2008ha, Denicol:2008rj}.

\subsection{Dispersion relation for bulk viscosity}
We consider a perturbation around the static quantities $X_0$
\begin{equation}
X= X_0 + \delta \tilde{X}, \; 
\delta \tilde{X}=\delta X e^{i(\omega t-\boldsymbol{k}\cdot \boldsymbol{r})}, \label{eq:perturbation}
\end{equation}
where we choose five independent variables $X =(\varepsilon, u^x, u^y, u^z, \Pi)$. Here, we only consider the system in the local rest frame, i.e. $u^\mu_0=(1,\boldsymbol{0})$. 
Then, we linearise eq.~\eqref{eq:energy_conservation},~\eqref{eq:ISshortbulk}
in vanishing magnetic fields and shear viscous tensor limit and obtain a cubic polynomial equation of the form given in eq.~\eqref{eq:ploy3rd} with $\mathsf{X}_i$'s are
\begin{align}
\mathsf{X}_0&=\frac{i}{\tau_{\Pi}}\alpha k^2, & \mathsf{X}_1&=-  \left(\alpha + \frac{1}{b_{1}} \right)k^2, & \mathsf{X}_2& =-\frac{i}{\tau_{\Pi}},
\end{align}
%\begin{equation}
%\omega ^3-\frac{i}{\tau _{\Pi }} \omega ^2-  \left(\alpha + \frac{1}{b_{1}} \right)k^2 \omega +\frac{i \alpha  }{\tau _{\Pi }}k^2=0,
%\label{eq:disonlybulk}
%\end{equation}
and the other two roots being zero. The solutions of this cubic polynomial are obtained from eq.~\eqref{eq:3rdployroot1}. Here, we introduce a constant $\alpha=c_s^2$, where $c_s$ is speed of sound.\par
We adopt the following parametrisation of the bulk viscosity coefficient and the relaxation time refs.~\cite{Denicol:2008ha, Denicol:2008rj}:
\begin{eqnarray}
\zeta &=&a_{1} s, \\
\tau_{\Pi} &=&\frac{\zeta}{\varepsilon+P} b_1=\frac{a_{1} b_{1}}{T},\label{Eq:b1}
\end{eqnarray}
where $s$ and $T$ are the entropy density and the temperature, respectively. The parameters $a_{1}$ and $b_{1}$ characterize the magnitudes of the viscosity and the relaxation time, respectively.

In the small wave-number limit, the dispersion relation is
\begin{eqnarray}
\omega=\left\{
\begin{array}{ll}
\frac{i}{\tau_{\Pi}}, \\ %-{\color{red} \frac{i\tau_\Pi }{b_1}k^2},\\\\
\pm k \sqrt{\alpha}.%+\frac{i\tau_{\Pi}}{2b_{1}}k^{2}.
\end{array}
\right.
\end{eqnarray}
Whereas the asymptotic forms of the dispersion relation in this case for large $k$ are
\begin{eqnarray}\label{eq:asybl}
\omega=\left\{
\begin{array}{ll}
i \frac{\alpha b_{1}}{\tau_{\Pi}(1+\alpha b_{1})}, \\ %+ \color{red}\mathcal{O}\left(\frac{1}{k^2}\right),\\\\
\pm k \sqrt{\alpha +\frac{1}{b_{1}}}+i\frac{1}{2 \tau_{\Pi}\left(1+\alpha b_{1}\right)}. %+ \color{red}\mathcal{O}\left(\frac{1}{k}\right) .
\end{array}
\right. 
\end{eqnarray}
Note that one of the roots is a pure imaginary which is also known as the non-hydrodynamic mode because it is independent of $k$ in the $k\rightarrow 0$ limit. From eq.~\eqref{eq:asybl} it is clear that the asymptotic group velocity is $v_{L}=\sqrt{\alpha +\frac{1}{b_{1}}}$. For the causal and stable propagation, the asymptotic group velocity must be subluminal i.e; $v_{L} \leq 1$ which imply $\frac{1}{b_{1}} \leq 1-\alpha$. For more details see ref.~\cite{Denicol:2008ha}.
\subsection{Dispersion relation for shear viscosity}
We use the following parametrization taken from ref.~\cite{Pu:2009fj} for the shear viscous coefficient and the corresponding relaxation time:
\begin{eqnarray}
\eta &=& a s,\\
\label{eq:defb}
\tau_{\pi} &=& \frac{\eta}{\varepsilon+P} b=\frac{a b}{T}.
\end{eqnarray}
Again we linearise eqs.~\eqref{eq:energy_conservation},~\eqref{eq:ISshortShear}~(the magnetic field and the bulk viscous pressure are taken to be zero) and obtain a set of equations with nine independent variables. Two of the roots are non-hydrodynamic with corresponding dispersion relation is $\omega=i/\tau_\pi$. Another four roots are
\begin{equation}
\omega=\frac{1}{2 \tau_{\pi}}\left(i \pm \sqrt{\frac{4 \eta \tau_{\pi}}{\varepsilon_{0}+P_{0}} k^{2}-1}\right),
\label{eq:SiFShMo}
\end{equation}
where each roots are double degenerate, they are known as the shear modes. The remaining three modes are obtained from a cubic polynomial of the form given in eq.~\eqref{eq:ploy3rd} with $X_i$'s are 
\begin{align}
X_0&=\frac{i}{\tau_{\pi}}\alpha k^2, & X_1&=-  \left(\alpha + \frac{4}{3b} \right)k^2, & X_2& =-\frac{i}{\tau_{\pi}}.
\end{align}
 These modes called sound modes as given in ref.~\cite{Pu:2009fj}. In the small $k$ limit, the dispersion relation for the sound modes are
\begin{equation}
\omega=\left\{\begin{array}{l}
{\frac{i}{\tau_{\pi}}}, \\
{\pm k \sqrt{\alpha}}.%+i \frac{2\tau_{\pi}}{3b} k^{2}
\end{array}\right.
\end{equation}
And in the large $k$ limit, the dispersion relations are
\begin{equation}\label{eq:smolak}
\omega=\left\{\begin{array}{l}
i\frac{3\alpha b}{\tau_{\pi}\left(4 + 3\alpha b\right)},\\
{\pm k \sqrt{\alpha+\frac{4}{3b}}+i\frac{2}{\tau_{\pi} (4 + 3\alpha b)}}.
\end{array}\right.
\end{equation}
The dispersion relations for the shear modes are given in eq.~\eqref{eq:SiFShMo} and the corresponding asymptotic group velocity is $v_{L}=\frac{1}{\sqrt{b}}$. So, for the causal and stable propagation of shear modes the condition $\frac{1}{b}\leq 1$, must hold. On the other hand, for sound modes, the dispersion relations in the large $k$ limit given in eq.~\eqref{eq:smolak} and the corresponding asymptotic group velocity is $v_{L}=\sqrt{\alpha +\frac{4}{3b}}$. So, the causality condition for sound modes are $\frac{1}{b} \leq \frac{3}{4}\left(1-\alpha\right)$. For more details see ref.~\cite{Pu:2009fj}.
%%%%%%%%%%%%%%%%%%%%%%%%%%%%%%%%%%%%%%%%%%%%%%%%%%
\section{Dispersion relation in the presence of magnetic field}
%\section{Stability analysis for MHD}
\label{sec4:ResultDis}

We extend our studies to explore the cases in a non-vanishing magnetic field. In this section, we will investigate the dispersion relation and the speed of sound in a viscous fluid in the presence of a homogeneous magnetic field. We will derive the physical conditions of causality and stability. To achieve this goal, we carry out a systematic study for the following cases,  (i) non-resistive ideal MHD, (ii) viscous MHD with bulk viscosity only, (iii) with shear viscosity only, (iv) with both bulk and shear viscosity.
%%%%%%%%%%%%%%%%%%%%%%%%%%%%%%%%%%%%%%%%%%%%%%%%%%%%%
%%%%%%%%%%%%%%%.  Ideal Fluid
\subsection{Ideal MHD}
\label{sec:ideal}
For an ideal non-resistive fluid in magnetic field the energy-momentum tensor eq.~\eqref{eq:EMTensFull} takes the following form
\begin{equation}
T^{\mu \nu}=\left(\varepsilon+P+B^{2}\right) u^{\mu} u^{\nu}-\left(P+\frac{B^{2}}{2}\right) g^{\mu \nu}-B^{2}b^{\mu} b^{\nu}.
\end{equation}
Here, we define 
\begin{equation}
b^{\mu} \equiv \frac{B^{\mu}}{B},
\end{equation}
which is normalized to $b^{\mu}b_{\mu}=-1$ and orthogonal to $u^{\mu}$ i.e, $b^{\mu}u_{\mu}=0$. 

Again we consider the similar perturbation as eq.~\eqref{eq:perturbation} around the equilibrium configuration in the local rest frame ($u^\mu_0=(1,\boldsymbol{0})$). 
Ignoring the second and higher-order terms for the perturbations in $\varepsilon, P, u^{\mu}$ and $B^{\mu}$, the perturbed energy-momentum tensor can be expressed as
\begin{eqnarray}
\delta{\tilde{T}}^{\mu\nu}&=\left(\varepsilon_{0}+P_{0}+B_{0}^{2}\right)\left(u_{0}^{\mu}\delta{\tilde{u}}^{\nu}+\delta{\tilde{u}}^{\mu}u_{0}^{\nu}\right)+\left(\delta{\tilde{\varepsilon}}+\delta{\tilde{P}}+2B_{0}\delta{\tilde{B}}\right)u_{0}^{\mu} u_{0}^{\nu}\notag\\
&-\left(\delta{\tilde{P}}+B_{0}\delta{\tilde{B}}\right)g^{\mu\nu}-B_{0}^{2}\left(b_{0}^{\mu}\delta{\tilde{b}}^{\nu}+\delta{\tilde{b}}^{\mu}b_{0}^{\nu}\right)-2B_{0}\delta{\tilde{B}}b_{0}^{\mu}b_{0}^{\nu}.
\end{eqnarray}
Next, using the above $\delta{\tilde{T}}^{\mu\nu}$ in the energy-momentum conservation equations and noting that $\partial_{\mu}\delta{\tilde{T}}^{\mu\nu}=0$ we get the following four equations
\begin{eqnarray}
\label{eq:linIdealEMCons1}
i\omega \delta{\tilde{\varepsilon}}-i k_{x}h\delta{\tilde{u}^{x}}-i k_{y}h\delta{\tilde{u}^{y}}-i k_{z}h\delta{\tilde{u}^{z}}+i k_{z}B_{0}^{2}\delta{\tilde{b}^{t}}+i\omega B_{0}\delta{\tilde{B}}&=&0,\\
-i k_{x} \alpha \delta{\tilde{\varepsilon}}+i \omega h \delta{\tilde{u}^{x}}+i k_{z}B_{0}^{2}\delta{\tilde{b}^{x}}-i k_{x}B_{0}\delta{\tilde{B}}&=&0,\\
-i k_{y} \alpha \delta{\tilde{\varepsilon}}+i \omega h \delta{\tilde{u}^{y}}+i k_{z}B_{0}^{2}\delta{\tilde{b}^{y}}-i k_{y}B_{0}\delta{\tilde{B}}&=&0,\\
\label{eq:linIdealEMCons4}
-i k_{z} \alpha \delta{\tilde{\varepsilon}}+i \omega h \delta{\tilde{u}^{z}}-i \omega B_{0}^{2} \delta{\tilde{b}^{t}}+i k_{x}B_{0}^{2}\delta{\tilde{b}^{x}}+i k_{y}B_{0}^{2}\delta{\tilde{b}^{y}}+i k_{z}B_{0}\delta{\tilde{B}}&=&0.
\end{eqnarray}
Here, we define \(h=\varepsilon_{0}+P_{0}+B_{0}^{2}\), and use $\delta{\tilde{P}}=\alpha \delta{\tilde{\varepsilon}}$. 
The relevant Maxwell's equations which govern the evolution of magnetic fields in the fluid is  $\epsilon^{\mu\nu\alpha\beta}\partial_{\beta}F_{\nu\alpha}=0$, which can also be written in the following form 
\begin{eqnarray}
\partial_{\nu}(B^{\mu}u^{\nu}-B^{\nu}u^{\mu})=0.
\end{eqnarray}
Linearizing the above Maxwell's equations lead to the following set of equations
\begin{eqnarray}
\label{eq:linMaxwell1}
i k_{x}B_{0}\delta{\tilde{b}^{x}}+i k_{y}B_{0}\delta{\tilde{b}^{y}}+i k_{z}\delta{\tilde{B}}&=&0,\\
\label{eq:linMaxwell2}
i k_{z}B_{0} \delta{\tilde{u}^{x}}+i \omega B_{0} \delta{\tilde{b}^{x}}&=&0,\\
i k_{z}B_{0} \delta{\tilde{u}^{y}}+i \omega B_{0} \delta{\tilde{b}^{y}}&=&0,\\
\label{eq:linMaxwell4}
-i k_{x}B_{0}\delta{\tilde{u}^{x}}-i k_{y}B_{0}\delta{\tilde{u}^{y}}+i \omega \delta{\tilde{B}}&=&0.
\end{eqnarray}
 %The degrees of freedom of the above set of equations are the $u^i$, $b^i$ with $i=x,y,z$ and two scalars which are magnitude of the magnetic field $B$ and energy density $\varepsilon$, leading to seven degrees of freedom. 
 The equations of motion are the energy-momentum conservation equations [Eqs.~\eqref{eq:linIdealEMCons1}-\eqref{eq:linIdealEMCons4}] and the  Maxwell's equations [eq.~\eqref{eq:linMaxwell1}-\eqref{eq:linMaxwell4}]. 
 However, we notice that eq.~\eqref{eq:linMaxwell1} does not include a time-derivative and it is a constraint equation for 
 $\delta \tilde{B}$, $\delta \tilde{b}^x$ and $\delta \tilde{b}^y$. This constraint is consistently propagated to the remaining  system of equations of motion. After replacing $\delta \tilde{B}$ by $\delta \tilde{b}^x$ and $\delta \tilde{b}^y$, 
 these equations become
\begin{equation}
A \delta \tilde{X}^T=0,
\end{equation}
where 
\begin{equation}\label{Eq:IdDof}
 \delta\tilde{X}=\left(\delta{\tilde{\varepsilon}},\delta{\tilde{u}}^{x},\delta{\tilde{u}}^{y},\delta{\tilde{u}}^{z},\delta{\tilde{b}}^{x},\delta{\tilde{b}}^{y}\right),
\end{equation}
and $A$ is a $6\times 6$ matrix of the following form
\begin{eqnarray}\label{eq:A_SF}
A=\left(
\begin{array}{cccccc}
 i \omega  & -i k_{x} h & -i k_{y} h & -i k_{z}\left(\varepsilon_{0}+P_{0}\right) & -i \frac{k_{x}}{k_{z}}\omega B_{0}^2 & -i \frac{k_{y}}{k_{z}}\omega B_{0}^2 \\
 -i \alpha  k_{x} & i \omega h  & 0 & 0 & i k_{z} B_{0}^2 \left(\frac{k_{x}^2+k_{z}^2}{k_{z}^2}\right)  & i\frac{k_{x} k_{y}}{k_{z}}B_{0}^2 \\
 -i \alpha  k_y & 0 & i  \omega  h & 0 & i\frac{ k_{x} k_{y}}{k_{z}}B_{0}^2 & i k_{z} B_{0}^2 \left(\frac{k_{y}^2+k_{z}^2}{k_{z}^2}\right) \\
 -i \alpha  k_z & 0 & 0 & i \omega \left(\varepsilon_{0}+P_{0}\right) & 0 & 0 \\
 0 & i  k_z B_0 & 0 & 0 & i \omega  B_0 & 0 \\
 0 & 0 & i  k_z B_0 & 0 & 0 & i \omega  B_0 \\
\end{array}
\right).
\end{eqnarray}
In deriving the above equations, we have also used the following condition $\delta{\tilde{u}}_{\mu} b^{\mu}+u_{\mu}\delta{\tilde{b}}^{\mu}=0$, for changing the variable from $\delta{\tilde{b}}^{t}$ to $\delta{\tilde{u}}^{z}$. 

Without loss of generality, we consider the magnetic field $b^\mu$ along the $z$-axis
and $k^\mu$ lies in the $x\mbox{-}z$ plane and making an angle $\theta$ with the magnetic field, i.e.,
\begin{eqnarray}
b^{\mu}_0&=&(0,0,0,1), \nonumber \\
k^{\mu}&=&(\omega,k\sin{\theta},0,k\cos{\theta}).
\end{eqnarray}
The dispersion relations are obtained by solving
\begin{equation}
\det(A)=0,
\end{equation}
which gives us six hydrodynamic modes. Two of these modes are the called Alfv\'en modes whose dispersion relation are given as
\begin{eqnarray}
\omega=\pm kv_{A}\cos{\theta}, \; v_{A}^2=\frac{B_{0}^2}{h},
\end{eqnarray}
where $v_{A}$ is the speed of Alfv\'en wave. The fluid displacement is perpendicular to the background magnetic field in this case and the Alfv\'en modes can be thought of as the usual vibrational modes that travel down a stretched string.\par 

The rest four modes correspond to the magneto-sonic modes with the following dispersion relations
\begin{eqnarray}
\omega = \pm v_{M}k,
\end{eqnarray}
where $v_{M}$ is the speed of the magneto-sonic waves
\begin{eqnarray}
v_{M}^2=\frac{1}{2}\left[ v_{A}^{2} + \alpha \left(1-v_{A}^{2}\sin^{2}{\theta}\right) \pm \sqrt{\left\{ v_{A}^{2} + \alpha \left(1-v_{A}^{2}\sin^{2}{\theta}\right) \right\} ^2 - 4 \alpha v_{A}^{2} \cos^{2}{\theta}}\right].
\label{eq:MSwave}
\end{eqnarray}
The $\pm$ sign before the square-root term is for the ``fast'' and the ``slow'' magneto-sonic waves, respectively. For $\theta=0$ we have two cases (i) when $v_{A}>\sqrt{\alpha}$ i.e, the velocity of Alfv\'en wave is faster than the sound wave, then the fast branch turns to be Alfv\'en type and the slow branch becomes sound type (ii) when $v_{A}<\sqrt{\alpha}$, then the fast and the slow branch becomes sound and Alfv\'en type, respectively. Whereas for $\theta=\frac{\pi}{2}$, the velocity of the slow magneto-sonic mode becomes zero and the velocity of the fast magneto-sonic wave is 
\begin{equation}
v_f^2=v_{A}^2 + \alpha \left(1 - v_{A}^2\right).\label{fastMsonicwave}
\end{equation}
More discussions can be found in refs.~\cite{Grozdanov:2016tdf, Hernandez:2017mch}.

%%%%%%%%%%%%%%%%%%%%%%%%%%%%%%%%%%%%%%%
%% BULK VISCOSITY
%%%%%%%%%%%%%%%%%%%%%%%%%%%%%%%%%%%%%%5
\subsection{MHD with bulk viscosity}
\label{subsec:bulkonly}
Next, we consider QGP with finite bulk viscosity and a non-zero magnetic field. Usually, the bulk viscosity is proportional to the interaction measure $(\varepsilon-3P)/T^{4}$ of the system and hence supposed to be zero for a conformal fluid. Lattice calculation as in refs.~\cite{Borsanyi:2013bia,Bazavov:2014pvz}
 shows that the interaction measure has a peak around the QGP to hadronic phase cross-over temperature $T_{\rm co}$. For the sake of simplicity, here we take $\zeta/s= \textrm{constant}$  in the following calculation. The energy-momentum tensor in this case takes the following form
\begin{equation}
T^{\mu \nu}=\left(\varepsilon+P+\Pi+B^2\right)u^\mu u^\nu-\left(P+\Pi+\frac{B^2}{2}\right)g^{\mu\nu}-B^{2}b^{\mu}b^{\nu}.
\label{eq:EMTens}
\end{equation}
As before, we can decompose the energy-momentum tensor into two parts: an equilibrium  and a perturbation around the equilibrium i.e.,
\begin{eqnarray}
T^{\mu\nu}=T_{0}^{\mu\nu}+\delta{\tilde{T}}^{\mu\nu}.
\label{eq:emdecom}
\end{eqnarray}
Here, the perturbed energy-momentum tensor takes the following form
\begin{eqnarray}
\delta{\tilde{T}}^{\mu\nu}&=\left(\varepsilon_{0}+P_{0}+B_{0}^{2}\right)\left(u_{0}^{\mu}\delta{\tilde{u}}^{\nu}+\delta{\tilde{u}}^{\mu}u_{0}^{\nu}\right)+\left(\delta{\tilde{\varepsilon}}+\delta{\tilde{P}}+\delta{\tilde{\Pi}}+2B_{0}\delta{\tilde{B}}\right)u_{0}^{\mu} u_{0}^{\nu}\notag\\
&-\left(\delta{\tilde{P}}+\delta{\tilde{\Pi}}+B_{0}\delta{\tilde{B}}\right)g^{\mu\nu}-B_{0}^{2}\left(b_{0}^{\mu}\delta{\tilde{b}}^{\nu}+\delta{\tilde{b}}^{\mu}b_{0}^{\nu}\right)-2B_{0}\delta{\tilde{B}}b_{0}^{\mu}b_{0}^{\nu}.
\label{eq:perT}
\end{eqnarray}
We choose the independent variables as
\begin{equation}
\delta \tilde{X}=\left(\delta \tilde{\varepsilon}, \delta \tilde{u}^{x}, \delta \tilde{u}^{y}, \delta \tilde{u}^{z}, \delta \tilde{b}^{x}, \delta \tilde{b}^{y}, \delta \tilde{\Pi}\right).
\end{equation}
These conservation equations can be cast into the form $A \delta \tilde{X}^T=0$ and setting $\det{A}=0$, we get
\begin{eqnarray}
\label{eq:AlfBulk}
\omega^2-v_A^2 k^2\cos^2{\theta}=0,\\
\label{eq:MSBulk}
\omega^5 +\mathsf{X}_4 \omega^4 + \mathsf{X}_{3} \omega^3 + \mathsf{X}_{2} \omega^2 + \mathsf{X}_{1} \omega + \mathsf{X}_{0}=0,
\end{eqnarray} 
where 
\begin{align}
\mathsf{X}_{0}&=-\frac{i}{\tau_{\Pi}}\alpha v_{A}^2  k^4\cos^{2}{\theta}, && \mathsf{X}_{1}=\left(\alpha + \frac{1}{b_1}\right)v_A^2 k^4 \cos^2{\theta},\nonumber\\
\mathsf{X}_{2}&=\frac{i}{\tau_{\Pi}}\mathsf{Y}k^2 , &&\mathsf{X}_{3}=-\left(\mathsf{Y} +\frac{1}{b_1}\left(1-v_A^2\sin^2{\theta}\right)\right)k^2,\nonumber\\
\mathsf{X}_4&=-\frac{i}{\tau_{\Pi}}, &&\mathsf{Y}=v_{A}^{2} + \alpha \left(1-v_{A}^{2}\sin^{2}{\theta}\right).
\end{align}
Here the term $b_1$ of eq.~\eqref{Eq:b1}, in the above equations can be recast into $b_1 \equiv\frac{h\tau_\Pi}{\zeta}(1-v_A^2)$. The solution of eq.~\eqref{eq:AlfBulk} gives the following dispersion relation
\begin{eqnarray}
\omega&=&\pm v_A k\cos{\theta}.
\label{eq:alfven}
\end{eqnarray}
These two solutions of eq.~\eqref{eq:alfven} correspond to the Alfv\'en modes where $v_A$ is the  Alfv\'en velocity. The rest five modes obtained from eq.~\eqref{eq:MSBulk} correspond to the magneto-sonic modes.  Generally, quintic equations cannot be solved algebraically. Fortunately,  we find solutions for some special cases discussed below.

For $\theta=0$, we find that two modes coincides with the Alfv\'en modes in eq.~\eqref{eq:alfven} and the remaining three modes are obtained from a third-order polynomial of the form given in eq.~\eqref{eq:ploy3rd},  with the coefficients $X_0, X_1, X_2$ given as
\begin{align}\label{eq:X_iBlkt0}
X_0&=\frac{i}{\tau_{\Pi}}\alpha k^2, & X_1&=-\left(\alpha+\frac{1}{b_1}\right)k^2, &X_2=-\frac{i}{\tau_{\Pi}}.
\end{align}
The solutions of this cubic polynomial can be written as
\begin{equation}
 \omega_l=\frac{1}{3} \left(  
 -\frac{\xi^{-(l-1)}\Delta_0}{C} - \xi^{(l-1)} C -X_2
 \right)
\label{eq:3rdpolyt0}
\end{equation}
where $l=1,2,3$, $\xi$ is the primitive cubic root of unity, i.e., $\xi=\frac{-1+\sqrt{-3}}{2}$  and the other variables $C,\Delta_0$ etc. are given in eq.~\eqref{Eq:C}.\par
For $\theta=\pi/2$, the eq.~\eqref{eq:MSBulk} reduces to a third-order polynomial of the form eq.~\eqref{eq:ploy3rd}, where $X_i$'s are given as
\begin{align}
X_0&=\frac{i}{\tau_{\Pi}}v_f^2k^2, & X_1&=-\left(v_f^2+\frac{\zeta}{h\tau_{\Pi}}\right)k^2, & X_2& =-\frac{i}{\tau_{\Pi}},
\end{align}
where $v_f$ is the group velocity for the fast magneto-sonic waves defined in eq.\eqref{fastMsonicwave} and the other two roots are zero.

\begin{figure}[tb]
\begin{center}
	\includegraphics[width=0.48\textwidth]{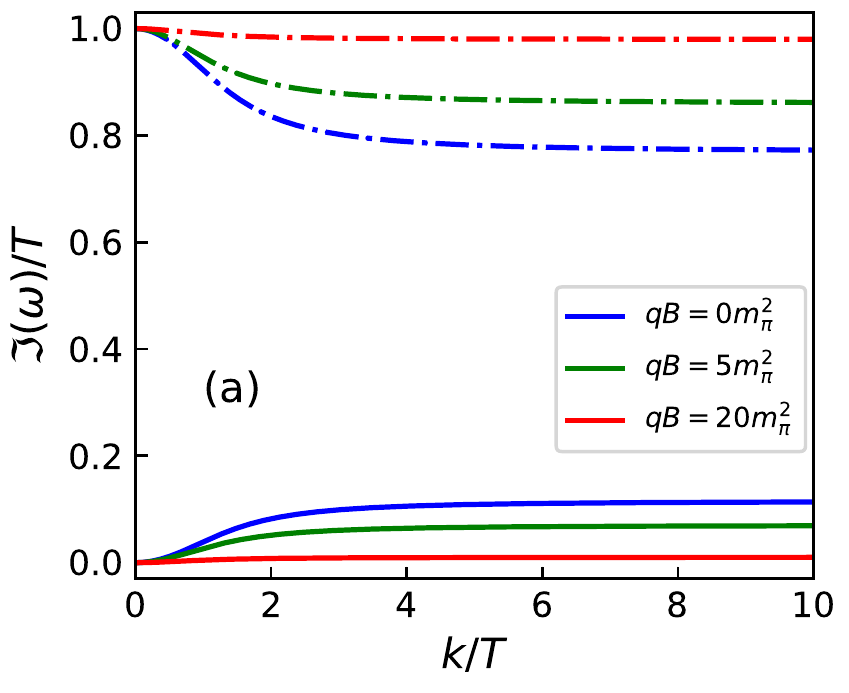}
	\includegraphics[width=0.48\textwidth]{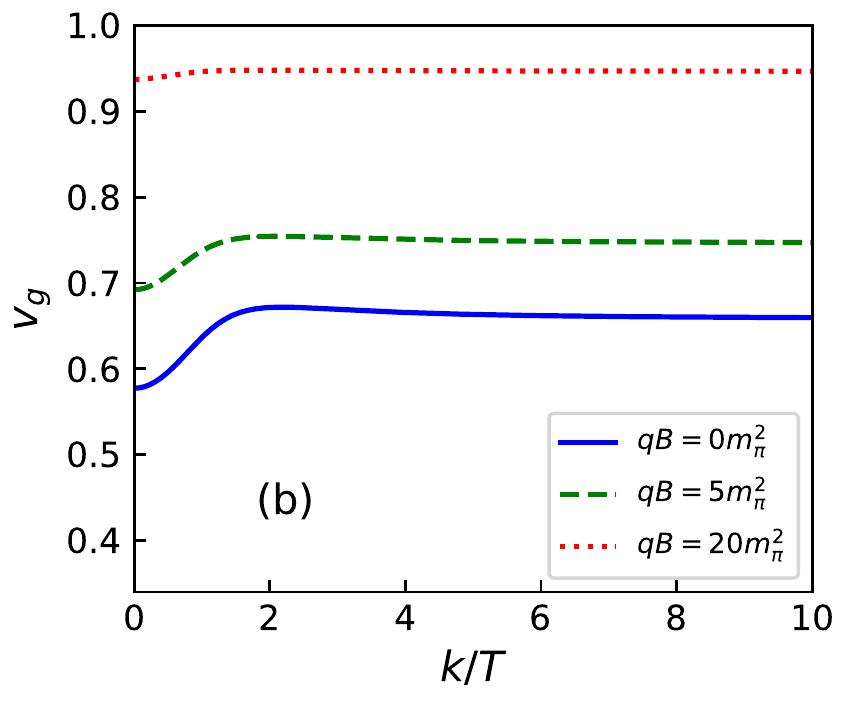}
\caption{(Color online) The imaginary parts of the dispersion relations obtain from eq.~\eqref{eq:MSBulk} for $\theta=\frac{\pi}{2}$ with different magnetic fields denoted by different colors. The blue, green and red colors correspond to  $B_{0}=$ 0, $5m_{\pi}^{2}$ and $20m_{\pi}^{2}$, respectively. In left panel the solid lines are for the propagating modes $(\omega_{2,3})$ and the dashed lines are for the non-propagating mode $(\omega_{1})$.  The other parameters used are $a_{1}=0.1$, $\alpha=1/3$, $T=200\,$MeV, $\tau_{\Pi}$=0.985 $\text{fm}$ and kept fixed for all the curves.
} 
	\label{fig:Pi/2MSBulk}
\end{center}	
\end{figure}

Note that all three roots in eq.~\eqref{eq:3rdpolyt0} are complex because the coefficients of eq.~\eqref{eq:X_iBlkt0} are complex and hence the phase velocity of any perturbations may contains a damping or growing and an oscillatory component. The left panel of Fig.~\ref{fig:Pi/2MSBulk} shows the imaginary part of the  normalised $\omega$  as a function of the  $k/T$ and the right panel shows the group velocity as a function of $k/T$ for different values of magnetic fields. Note that the imaginary part of the non-propagating mode increases and imaginary part of the propagating modes decreases when the magnetic field increases.
%, where we have chosen the parameters $a$=0.1, $T=200$ MeV, $\tau_{\Pi}=0.985$ fm$^{-1}$.
But it is clear that $\Im({\omega})$ always lies in the upper half of the complex plane for the parameters considered here\footnote{We have also checked the stability of the system by using Routh-Hurwitz stability criteria from the roots of eq.~\eqref{eq:X_iBlkt0}. We found that these roots always corresponds to stable states even for a general set of fluid parameters including the one used in the current work.}. This implies that any perturbation will always decay and the fluid is always stable. Also, for this parameter set-up the group velocity~$v_g\leq 1$, so the wave propagation is causal. 
%In Fig.~\ref{fig:Pi/2MSBulk} the blue, green and red colors correspond to  $B_{0}=$ 0, $5m_{\pi}^{2}$ and $20m_{\pi}^{2}$, respectively.

%\begin{figure}[h!tb]
\par
If we take the small $k$ limit, Eqs.~\eqref{eq:AlfBulk} and~\eqref{eq:MSBulk} yield the following modes:
\begin{eqnarray}
\omega=\left\{\begin{array}{l}
{\frac{i}{\tau_{\Pi}}},\\
\pm kv_A\cos{\theta} ,\\
{\pm k v_M}.
\end{array}\right.
\end{eqnarray}
For this case the group velocity is observed to be same as the velocity for the ideal MHD.
\begin{figure}[h!]
\begin{center}
	\includegraphics[width=0.48\textwidth]{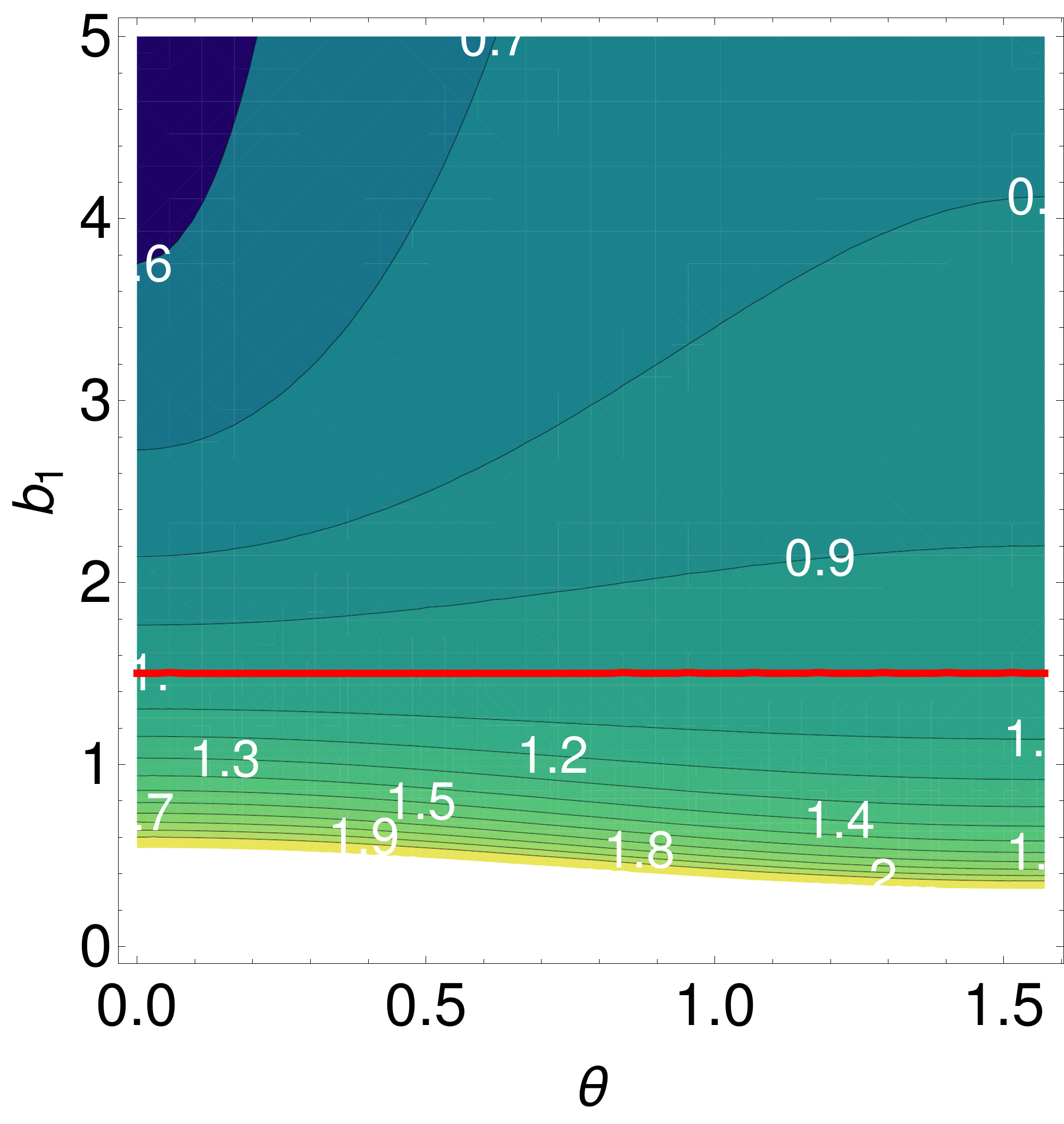}
	\includegraphics[width=0.48\textwidth]{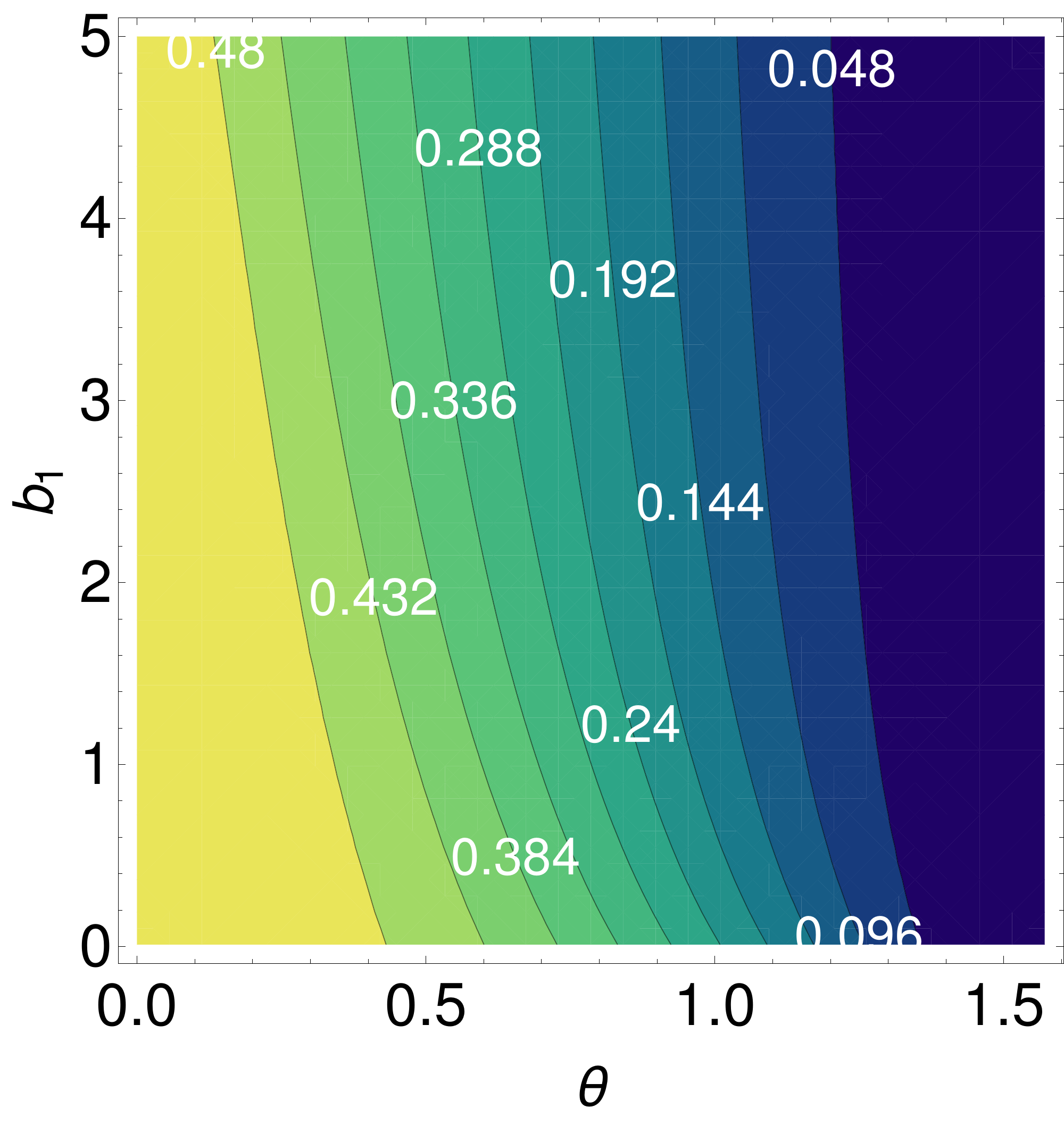}
\caption{(Color online) Contour plot showing various causal regions, obtained from eq.~\eqref{eq:vLbuk}, for fast (left panel) and slow (right panel) branches. The red contour is the critical line of causality, denoting $v_L^2=1$. The region above the red line is causal for the fast magneto-sonic waves and acausal below. The slow branch is causal throughout the parameter space. The magnitude of the magnetic field has been fixed to $qB=10m^2_{\pi}$ and the other parameters used are $\alpha=1/3$, $T=200\,$MeV.}
	\label{fig:conbulk}
\end{center}	
\end{figure}
%%%%%%%%%%%%%%%%%%%%%%%%%%%%

We analyse the causality of the system by following ref.~\cite{Pu:2009fj} where
 it was shown that to guarantee the causality requires that the asymptotic value of the group 
 velocity should be less than the speed of light. Alfv\'en mode in eq.~\eqref{eq:AlfBulk} remains 
 unaffected due to the bulk viscosity and hence always remain causal. For the magneto-sonic waves in the large $k$ limit, we take the following ansatz $\omega=v_L k$ in eq.~\eqref{eq:MSBulk} and collect terms in the leading-order of $k$, this yields
\begin{equation}
v_L^4 -x v_L^2 + y=0,
\label{eq:v4th}
\end{equation}
where
\begin{eqnarray}
&&x=v_{A}^2 + \left(\alpha+\frac{1}{b_1}\right) \left( 1- v_{A}^2\sin^{2}{\theta}\right),\nonumber\\
&&y=\left(\alpha+\frac{1}{b_1}\right) v_{A}^2\cos^2{\theta}.
\end{eqnarray}
The velocities $v_L$ are
\begin{eqnarray}
v_L^{2}=\frac{1}{2}\left(x\pm\sqrt{x^{2} -4 y}\right). \label{eq:vLbuk}
\end{eqnarray}
Here, we see that unlike the small $k$ limit, at large $k$ the  group velocity is affected by the transport coefficients. In order to have causal propagation, one demands $v_L^2 \leq 1$, which yields a causal parameter-set for the two branches, which correspond to the fast or slow magneto-sonic modes
\begin{eqnarray}\label{Eq:CausBulk}
\text{fast:~~}&&(0<y<1) \land (2\sqrt{y}\leq x<y+1),\nonumber\\
%\text{slow:~~}\alpha +\frac{\sin^{2}{\theta}}{b_{1}}\left(1-\alpha \cos^{2}{\theta}\right\leq 1,\\
\text{slow:~~}&&\left[\left(0<y<1\right)\land \left(x\geq 2\sqrt{y}\right)\right]\lor \left[(y\geq 1)\land (x>y+1)\right]
\end{eqnarray}
Contour plot of the various causal regions is shown in Fig.~\ref{fig:conbulk}, where $b_1$ is defined in eq.~\eqref{Eq:b1}. For the fast branch, we find that, although the asymptotic velocities depend on the magnitude of the magnetic field and the direction $\theta$, the critical value, i.e., $b_1=1.5$ (red solid line), is independent of them.
The slow branch is similarly $B$ and $\theta$ dependent but moreover is causal throughout the parameter space. 

%%%%%%%%%%%%%%%% Shear + Magnetic field %%%%%%%%%%%%%%%%%%%%%%%%%%%%%%%%%%%%%

\subsection{MHD with shear viscosity}
\label{sec:shear}
Many theoretical studies indicate that shear viscosity over entropy $\eta/s$ has a minimum near the crossover temperature $T_{co}$ and rises as a function of temperature on both sides of $T_{co}$ in ref.~\cite{Kovtun:2004de}. Although such studies indicate $\eta/s$ to be temperature dependent, nevertheless that would require an additional parametrization of $\eta/s$ which should come from the underlying theory. For simplicity, we will assume in the foregoing section $\eta/s$ is a constant.\par 
The energy-momentum tensor for a fluid with zero bulk and non-zero shear viscosity in a magnetic field takes the following form
\begin{eqnarray}
T^{\mu \nu}=\left(\varepsilon+P+B^{2}\right) u^{\mu} u^{\nu}-\left(P+\frac{B^{2}}{2}\right) g^{\mu \nu}-B^{2}b^{\mu} b^{\nu}+\pi^{\mu\nu}
\label{eq:shearT}.
\end{eqnarray}
According to the IS second-order theories of relativistic dissipative fluid dynamics, the space-time evolutions of the shear stress tensor are given by eq.~\eqref{eq:ISshortShear}.
For a given perturbation in the fluid,  the energy-momentum tensor and the shear stress tensor can be decomposed as
\begin{eqnarray}
T^{\mu\nu}&=&T_0^{\mu\nu}+\delta \tilde{T}^{\mu\nu} ,\\ 
\pi^{\mu\nu}&=&\pi_0^{\mu\nu}+\delta\tilde{\pi}^{\mu\nu}.
\label{eq:sheardecompose}
\end{eqnarray} 
Where the perturbed energy-momentum tensor is
\begin{eqnarray}
\delta{\tilde{T}}^{\mu\nu}&=\left(\varepsilon_{0}+P_{0}+B_{0}^{2}\right)\left(u_{0}^{\mu}\delta{\tilde{u}}^{\nu}+\delta{\tilde{u}}^{\mu}u_{0}^{\nu}\right)+\left(\delta{\tilde{\varepsilon}}+\delta{\tilde{P}}+2B_{0}\delta{\tilde{B}}\right)u_{0}^{\mu} u_{0}^{\nu}\notag\\
&-\left(\delta{\tilde{P}}+B_{0}\delta{\tilde{B}}\right)g^{\mu\nu}-B_{0}^{2}\left(b_{0}^{\mu}\delta{\tilde{b}}^{\nu}+\delta{\tilde{b}}^{\mu}b_{0}^{\nu}\right)-2B_{0}\delta{\tilde{B}}b_{0}^{\mu}b_{0}^{\nu}+\delta \tilde{\pi}^{\mu\nu}.
\label{eq:shearperturb}
\end{eqnarray}
As usual, to solve the set of equations  eq.~\eqref{eq:ISshortShear}, the conservation of the perturbed energy-momentum tensor~[eq.~\eqref{eq:shearperturb}], and eqs.~\eqref{eq:linMaxwell1}-\eqref{eq:linMaxwell4} for obtaining the dispersion relation we write them in a matrix form
\begin{eqnarray}
A\delta \tilde{X}^T=0,
\label{eq:shearAX}
\end{eqnarray}
where $\delta \tilde{X}=(\delta\tilde{\varepsilon},\delta\tilde{u}^{x},\delta\tilde{u}^{y},\delta\tilde{u}^{z},\delta\tilde{b}^{x}, \delta\tilde{b}^{y},\delta\tilde{\pi}^{xx},\delta\tilde{\pi}^{xy},\delta\tilde{\pi}^{xz}, \delta\tilde{\pi}^{yy},\delta\tilde{\pi}^{yz})$ and the matrix $A$  given in eq.~\eqref{eq:A_for_Sh}. The $\det(A)=0$ gives
\begin{eqnarray}
\label{eq:non-hSh}
\left(1 + i \omega \tau_{\pi}\right)^2&=&0,\\
\label{eq:AlfSh}
\omega^3 -\frac{i}{\tau_{\pi}}\omega^2 -\left(v_{A}^2\cos^2{\theta}+\frac{\eta}{h\tau_{\pi}}\right)k^2 \omega + \frac{i}{\tau_{\pi}} k^2 v_{A}^2\cos^2{\theta}&=&0,\\
\label{eq:MSShear}
\omega^6 + \mathsf{X}_{5}\omega^5 + \mathsf{X}_{4}\omega^4 + \mathsf{X}_{3}\omega^3 + \mathsf{X}_{2}\omega^2 + \mathsf{X}_{1}\omega + \mathsf{X}_{0}&=&0,
\end{eqnarray}
where
\begin{eqnarray}
\mathsf{X}_{5}&=&- \frac{2i}{\tau_\pi},\nonumber\\
\mathsf{X}_{4}&=&-\frac{1}{\tau_{\pi}^2} - \left[ \mathsf{Y} + \frac{1}{3b}\left\{ 7- v_{A}^{2}\left(3+\sin^{2}{\theta}\right)\right\} \right]  k^{2}, \nonumber\\
\mathsf{X}_{3}&=&\frac{i}{\tau_{\pi}}\left[2\mathsf{Y} + \frac{1}{3b}\left\{ 7- v_{A}^{2}\left(3+\sin^{2}{\theta}\right)\right\} \right]  k^{2}, \nonumber\\
\mathsf{X}_{2}&=& \frac{\mathsf{Y}}{\tau_{\pi}^2} k^2 + \left[ \alpha \left(v_{A}^2\cos^2{\theta}+\frac{\eta}{h\tau_{\pi}}\right) +\frac{1}{3b}\left\{ \frac{4\eta}{h\tau_{\pi}}+ v_A^2\left(3 + \sin^{2}{\theta}\right)\right\}\right]k^4, \nonumber\\
\mathsf{X}_{1}&=&-\frac{i}{\tau_{\pi}}\left[ \alpha \left(2v_{A}^2\cos^2{\theta}+\frac{\eta}{h\tau_{\pi}}\right) +\frac{v_{A}^2}{3b}\left(3+\cos^{2}{\theta}\right)\right] k^4, \nonumber\\
\mathsf{X}_{0}&=& - \frac{\alpha}{\tau_{\pi}^2} v_{A}^2 k^4\cos^{2}{\theta} , \hspace*{1cm}  \hspace*{1cm} \mathsf{Y}=  v_{A}^{2} + \alpha \left(1-v_{A}^{2}\sin^{2}{\theta} \right) .
\end{eqnarray}
%%%%%%%%%%%%%%%%%%%%%%%%
Note that the term $b$ of eq.~\eqref{eq:defb} in the above equations can be recast into $b\equiv\frac{h\tau_\pi}{\eta}(1-v_A^2)$. From eq.~\eqref{eq:non-hSh} we get two non-propagating and stable modes as
\begin{equation}
\omega=\frac{i}{\tau_{\pi}}.
\end{equation} 
%which is clearly a non-hydrodynamic mode and it's a doubly degenerate solution.\par
%%%%%%%%%%%%%%%%%%%%%%%%%%%%%%%%%%%%
Equation~\eqref{eq:AlfSh} is a third-order polynomial equation and the analytic solution for this type of equation is discussed in appendix~\ref{sec:ployeqns}. Equation~\eqref{eq:MSShear} is a sixth-order polynomial equation which is impossible to solve analytically. We can still gain some insight for a few special cases which are discussed below. 
\par
%%%%%%%%%%%%%%%%
For $\theta =0$, eq.~\eqref{eq:AlfSh} still remains a third-order polynomial equation and the coefficients of that polynomial can easily be obtained from eq.~\eqref{eq:AlfSh} as
\begin{align}
X_0&=\frac{i}{\tau_{\pi}} v_{A}^2 k^2, & X_1&=-\left(v_{A}^2+\frac{\eta}{h\tau_{\pi}}\right)k^2, & X_2& =-\frac{i}{\tau_{\pi}}.
\label{eq:X_iMSSAlf}
\end{align}	
%%%%%%%%%%%%%%%%	
On the other hand, eq.~\eqref{eq:MSShear} can be factorized into two third-order polynomial equations. The coefficients of one of such the third-order polynomial equation are
\begin{align}
X_0&=\frac{i  \alpha}{\tau_{\pi}} k^2, & X_1&=-\left(\alpha+\frac{4}{3 b}\right)k^2, & X_2& =-\frac{i}{\tau_{\pi}},
\label{eq:X_iMSSSou}
\end{align}
whereas the coefficients of the remaining other third-order polynomial equation from eq.~\eqref{eq:MSShear} are same as eq.~\eqref{eq:X_iMSSAlf}

The roots of these third order polynomial equations are discussed in appendix~\ref{sec:ployeqns} with the given $X_{i}$s.
We checked that the dispersion relations obtained from these equations with the coefficients given in eq.~\eqref{eq:X_iMSSSou} are same as the sound mode in ref.~\cite{Pu:2009fj}.\par For $\theta =\pi/2$, one root of eq.~\eqref{eq:AlfSh} vanish and other two roots are of the form
\begin{eqnarray}%\label{eq:Pi/2shear}
\omega=\frac{1}{2\tau_{\pi}}\left(i \pm \sqrt{\frac{4 \eta \tau_{\pi}}{h} k^{2} - 1}\right).
\end{eqnarray}
From eq.~\eqref{eq:MSShear}, one of the root vanish and other two roots are of the form
\begin{eqnarray}\label{eq:Pi/2shear}
\omega=\frac{1}{2\tau_{\pi}}\left(i \pm \sqrt{\frac{4 \eta \tau_{\pi}}{\varepsilon_{0}+P_{0}} k^{2} - 1}\right).
\end{eqnarray}
The remaining three modes from eq.~\eqref{eq:MSShear}, are obtained from a cubic polynomial with $X_i$'s given as:
\begin{align}
X_0&=\frac{i}{\tau_{\pi}}v_f^2 k^2, & X_1&=-\left( v_f^2 +\frac{4\eta}{3 h\tau_{\pi}}\right)k^2, & X_2& =-\frac{i}{\tau_{\pi}}.
\label{eq:X_iMSSFast}
\end{align}
The corresponding roots can be calculated using the formula given in appendix~\ref{sec:ployeqns}.\par 
\begin{figure}[ht]
\begin{center}
	\includegraphics[width=0.49\textwidth]{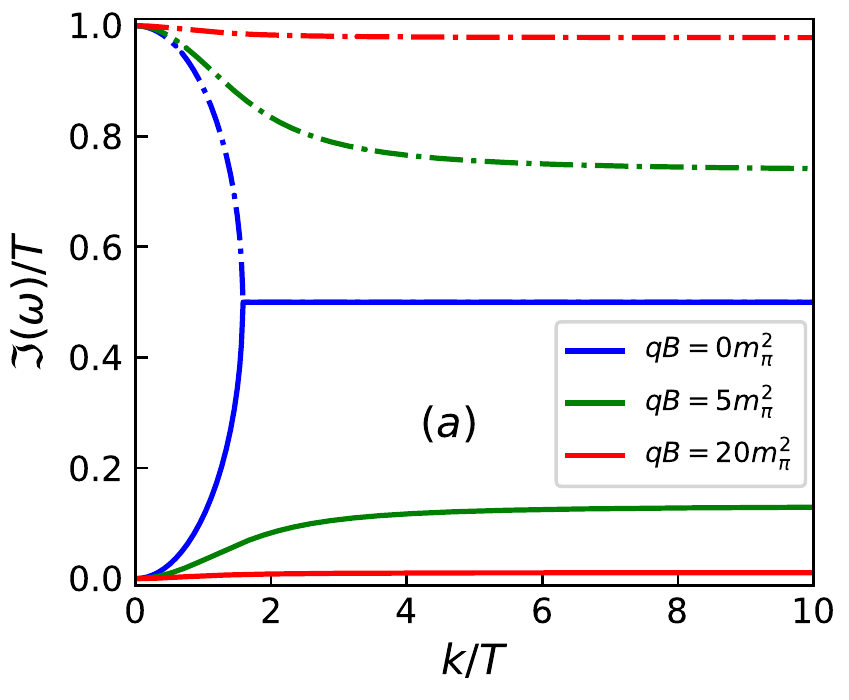}
	\includegraphics[width=0.49\textwidth]{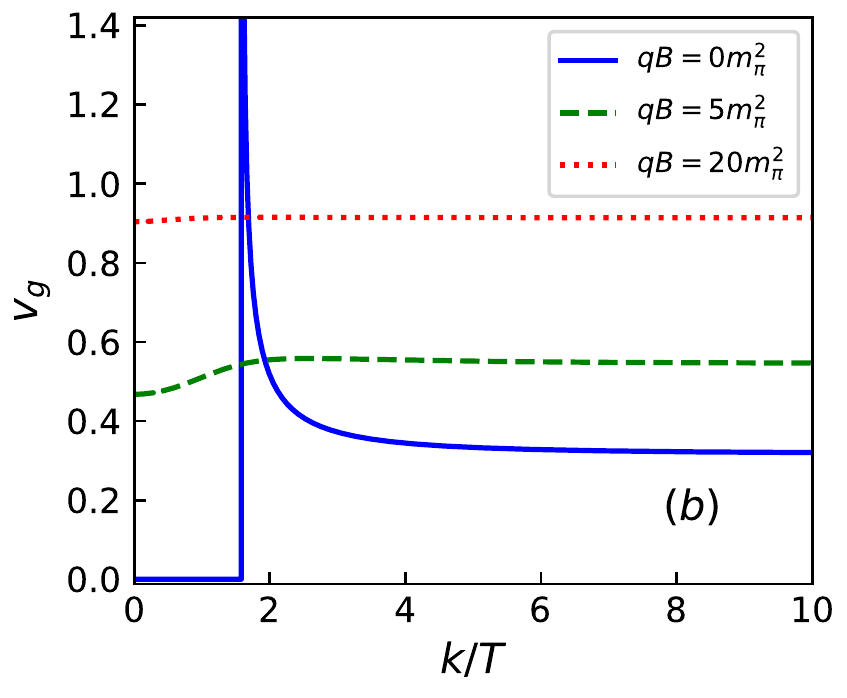}
	\caption{(Color online) The left panel shows the imaginary parts of the dispersion relations and the right panel shows the group velocities obtained from a cubic polynomial with the coefficients given in eq.~\eqref{eq:X_iMSSAlf} for $\theta=0$. The other parameters used are \(a=0.1, \alpha=1/3, T=200\, \text{MeV}, \tau_{\pi}= 0.985\, \text{fm}\) and their values are kept fixed for all the curves. In the left panel, the solid lines are for $\Im(\omega_{2,3})$ which are degenerate. The dash-dotted lines correspond to $\Im{(\omega_{1})}$.}
	\label{fig:AlfShear}
\end{center}
\end{figure}
\begin{figure}[t]
\begin{center}
	\includegraphics[width=0.49\textwidth]{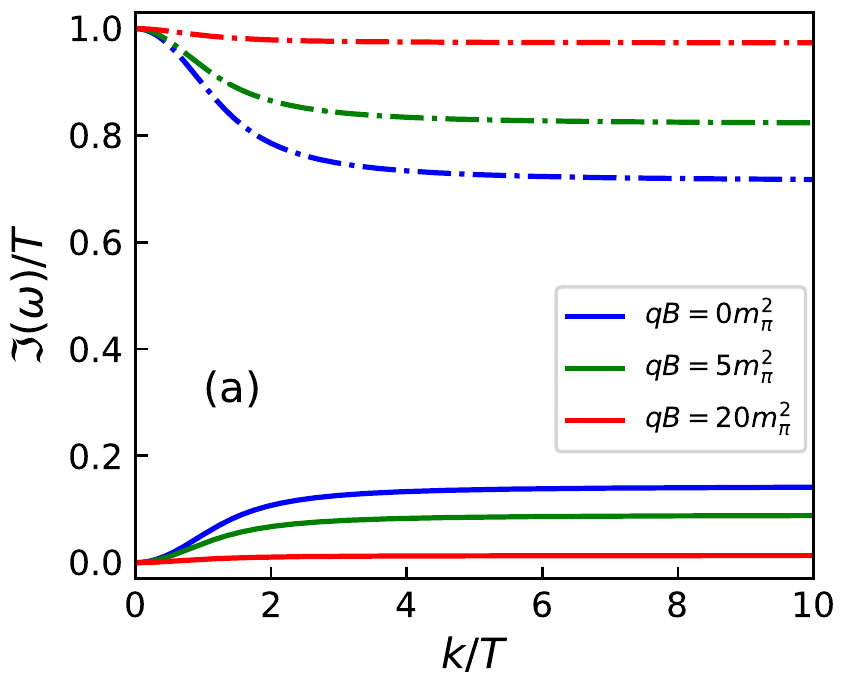}
	\includegraphics[width=0.49\textwidth]{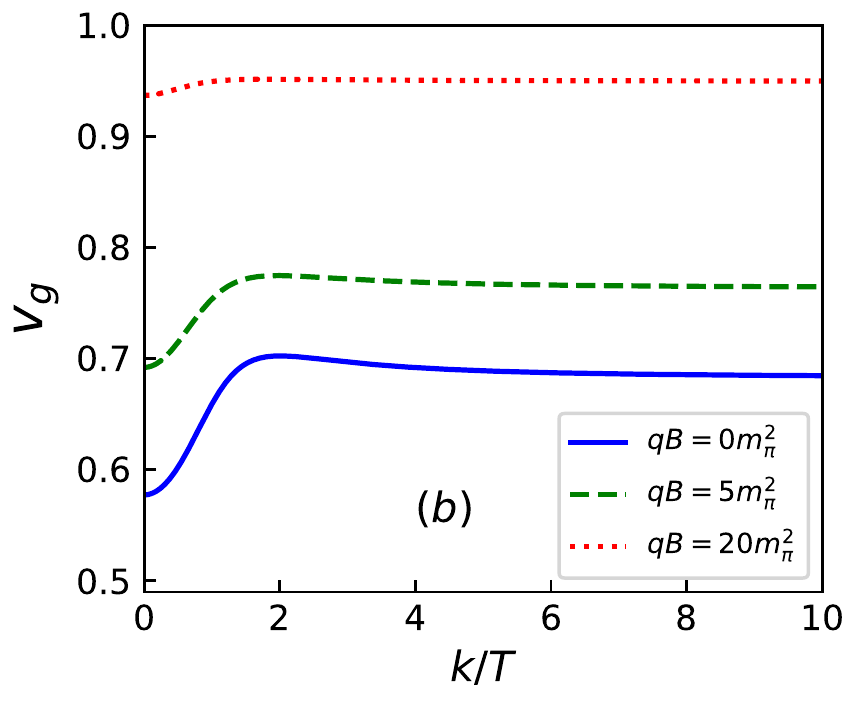}
	\caption{(Color online) The left panel shows the imaginary parts of $\omega$ and the right panel shows the group velocities obtained from the cubic polynomial with the coefficient given in eq.~\eqref{eq:X_iMSSFast} for $\theta=\frac{\pi}{2}$. The other parameters used are \(a=0.1, \alpha= 1/3,  T=200\, \text{MeV}, \tau_{\pi}=0.985\, \text{fm}\) and are kept fixed for all the curves.In the left panel, the dash-dotted lines represent $\Im{(\omega_{1})}$ and the solid lines are for $\Im(\omega_{2,3})$ which are also degenerate.
	}
	\label{fig:MSShear}
\end{center}	
\end{figure}

The left panel of Fig.~\ref{fig:AlfShear} shows the dependence of the imaginary parts of $\omega$ as a function of $k/T$ and the right panel shows the group velocity as a function of $k/T$ for different values of magnetic field for $\theta=0$. Various lines corresponds to different magnetic fields: $qB=0$ (blue lines), $qB=5m_{\pi}^{2}$ (green lines), $qB=20m_{\pi}^2$ (red lines). Fig.~\ref{fig:MSShear} shows the same thing but for $\theta=\frac{\pi}{2}$ (eq.~\eqref{eq:X_iMSSFast}). %Fig.~\ref{fig:AlfShear} the solid lines are for $\Im(\omega_{2,3})$ which are degenerate. The dash-dotted lines correspond to $\Im{(\omega_{1})}$.
\par
%In the left panel of Fig.~\ref{fig:MSShear} the dash-dotted lines represent $\Im{(\omega_{1})}$ and the solid lines are for $\Im(\omega_{2,3})$ which are also degenerate. \par

In the small $k$ limit the dispersion relations that we get from eqs.~\eqref{eq:non-hSh}-\eqref{eq:MSShear} are
\begin{eqnarray}
\omega=\left\{\
\begin{array}{ll}
\frac{i}{\tau_{\pi}},\\
%\frac{i}{\tau_{\pi}}-\frac{i\eta}{h}k^2+\mathcal{O}(k^3),\\
\pm k v_{A}\cos{\theta},%+\frac{i\eta}{2h}k^2+\mathcal{O}(k^3),
\\
%\frac{i}{\tau_{\pi}}+\mathcal{O}(k),\\
\pm k v_M. %+\mathcal{O}(k^2).
\end{array}
\right.
\end{eqnarray}
\begin{figure}
\begin{center}
	\includegraphics[width=0.48\textwidth]{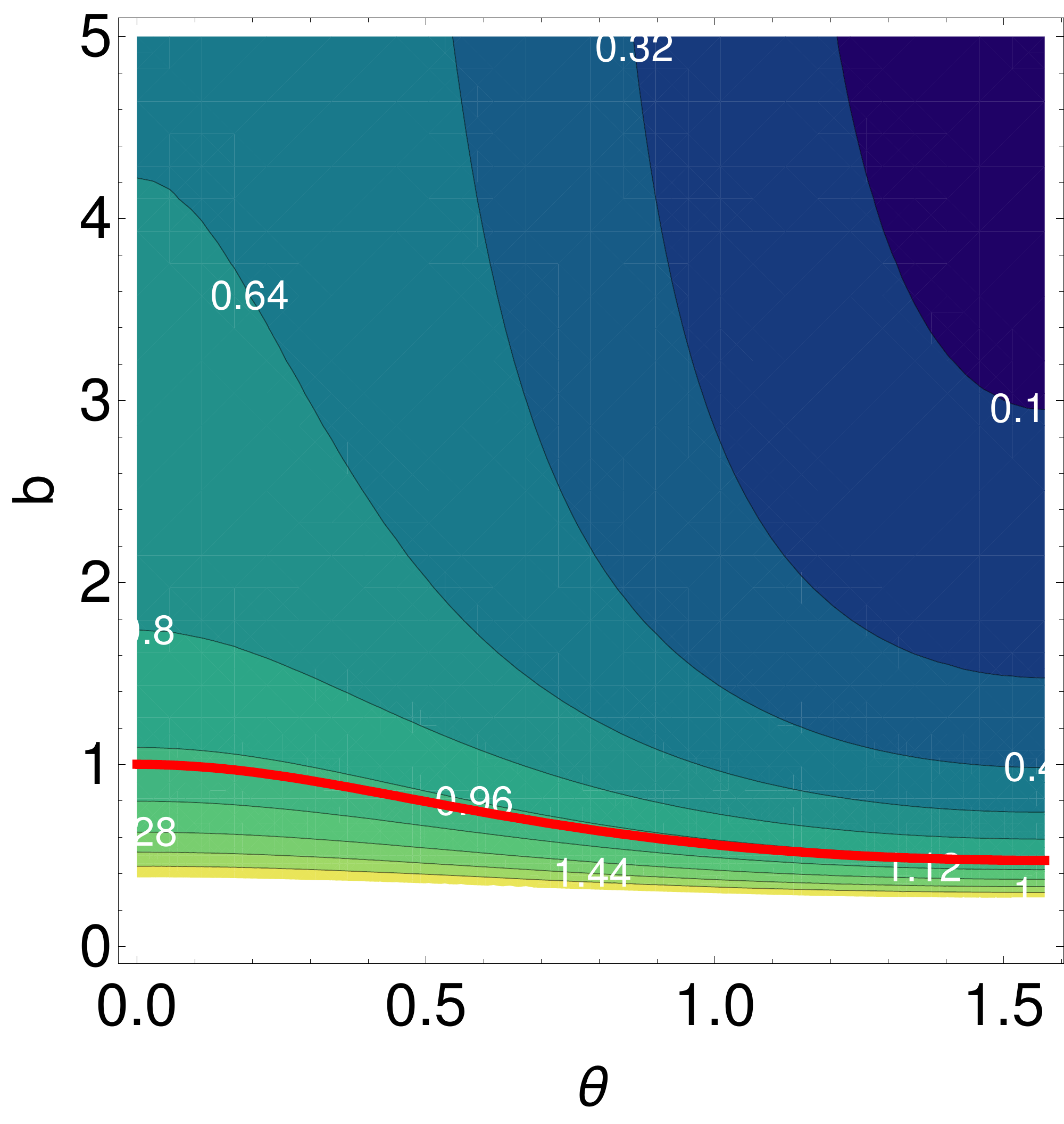}
	\includegraphics[width=0.48\textwidth]{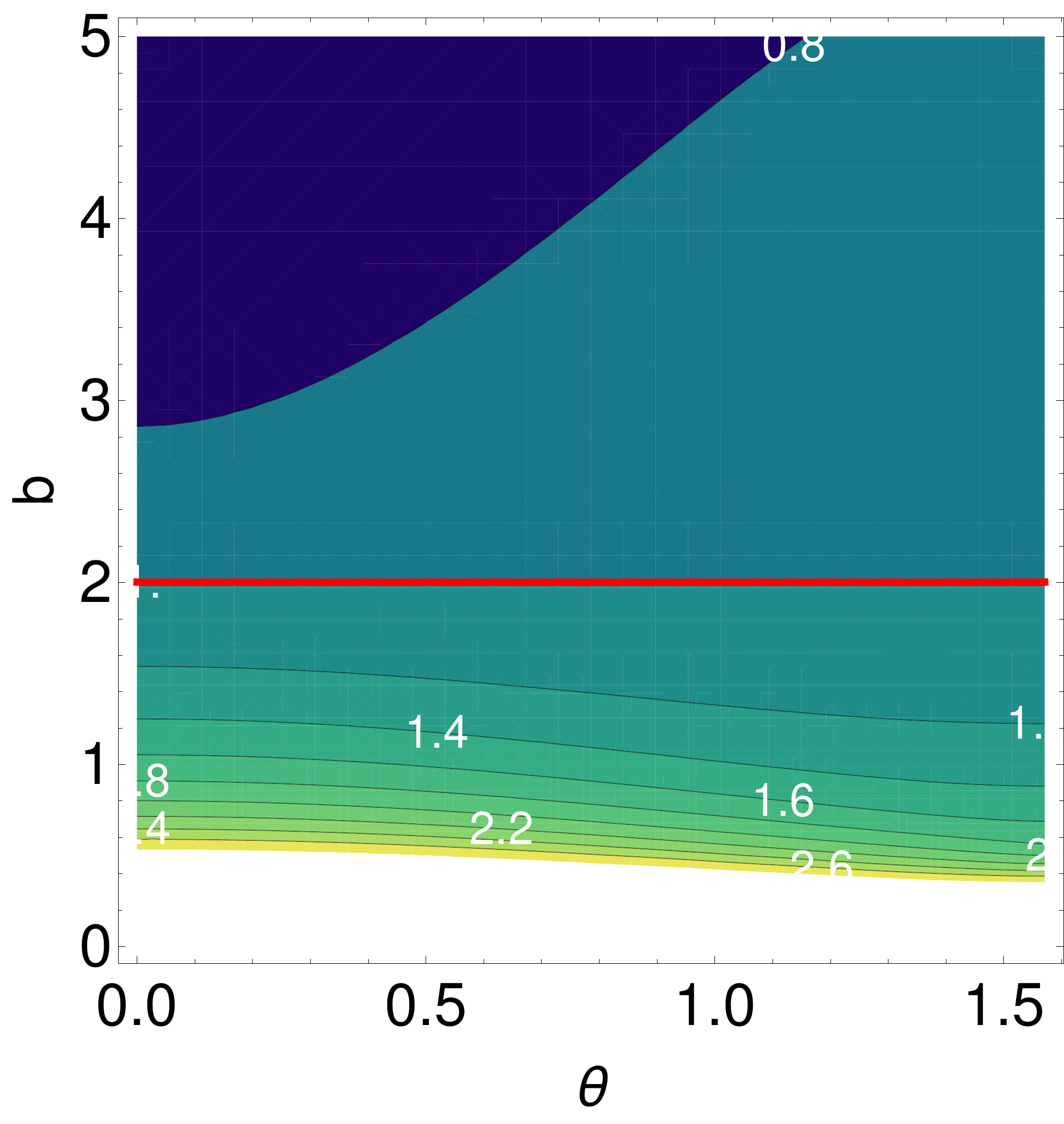}
    \includegraphics[width=0.48\textwidth]{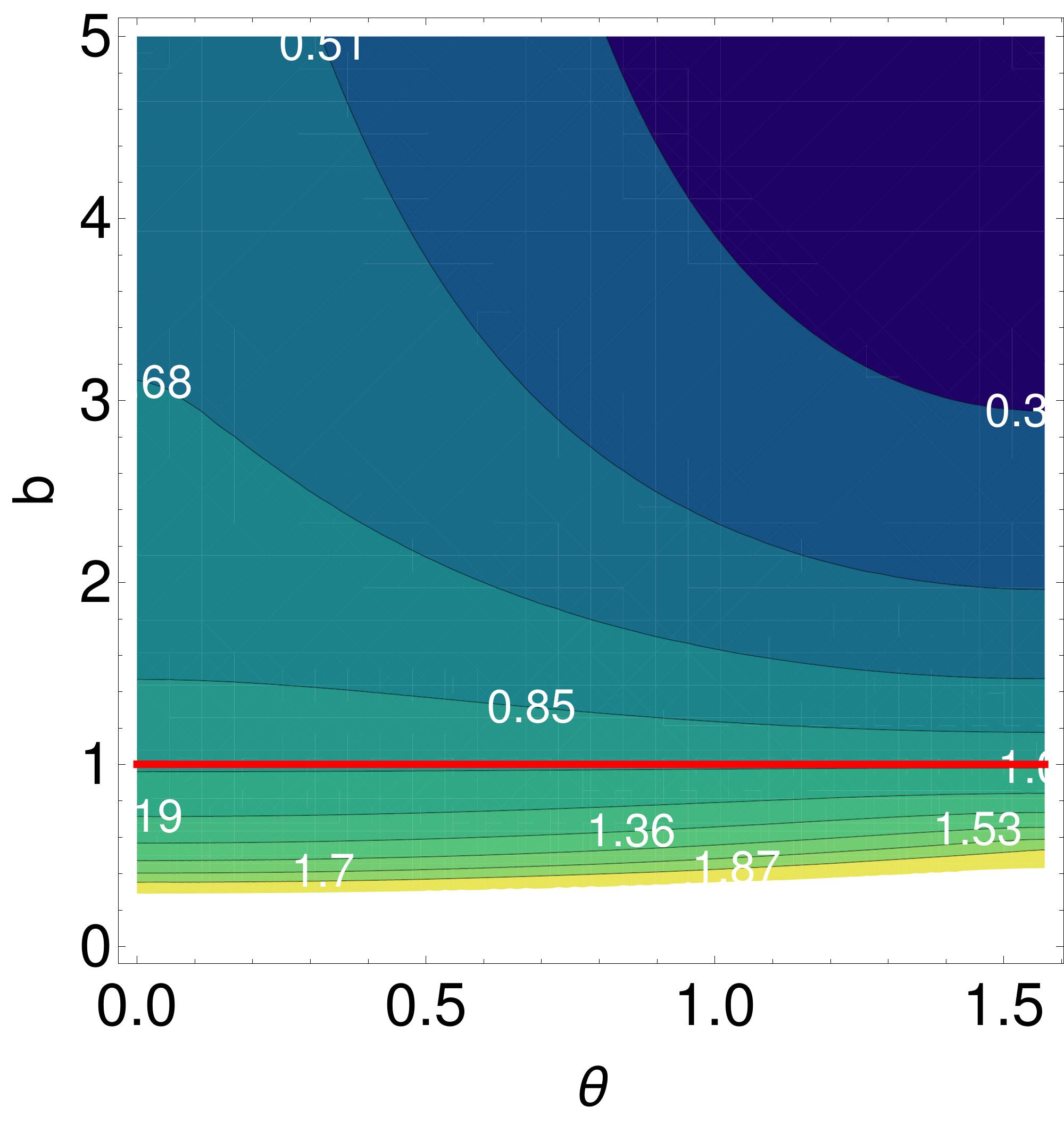}
\caption{(Color online) Contour plot showing various causal regions, obtained from eq.~\eqref{eq:v_LforSh}, for Alfv\'en mode (top left) and the set of fast (top right) and slow modes (bottom) from eq.~\eqref{eq:v_LforSh}. The red contour is the critical line of causality, denoting $v_L^2=1$. The region above the red line is causal and below it corresponds to the acausal zone. The magnitude of the magnetic field has been fixed to $qB=10m^2_{\pi}$ and the other parameters used are $\alpha=1/3$, $T=200\,$MeV.}
	\label{fig:conshear}
\end{center}
\end{figure}
Note that the first root have a degeneracy five.

In the large $k$ limit we use the ansatz $\omega=v_L k$ and keep only the leading-order terms in $k$, then the velocities $v_L$ are
\begin{eqnarray}\label{eq:v_LforSh}
v_L^{2}=\left\{\
\begin{array}{ll}
v_A^2\cos^2{\theta}+\frac{\eta}{h\tau_{\pi}},\\
\frac{1}{2}\left[x\pm\sqrt{x^{2} - 4 y}\right],
\end{array}
\right. \label{eq:vLshear}
\end{eqnarray}
where
\begin{eqnarray}
&&x=v_{A}^{2} + \alpha \left(1-v_{A}^{2}\sin^{2}{\theta}\right) +\frac{1}{3b}\left\{\ 7-v_{A}^{2}\left(3+\sin^{2}{\theta}\right)\right\}\ ,\nonumber\\
&&y=\alpha \left(v_{A}^{2}\cos^{2}{\theta}+\frac{\eta}{h \tau_{\pi}}\right)+\frac{1}{3b}\left\{\ v_{A}^{2}\left(3+\cos^{2}{\theta}\right)+\frac{4\eta}{h\tau_{\pi}}\right\}\ .
\end{eqnarray}
The asymptotic causality condition  for the shear-Alfv\'en mode can readily be obtained as
\begin{eqnarray}\label{Eq:StabRegSH}
&\text{shear-Alfv\'en:~~}&v_A^2\cos^{2}{\theta}+\frac{(\varepsilon+P)}{hb} \leq 1,
\end{eqnarray}
where $b$ is defined in eq.~\eqref{eq:defb}. We observe that in this case the wave velocity and the causality conditions depend on both the magnitude of the magnetic field and direction of propagation of the perturbation. To explore the inter-dependency we show various causal regions as a function of $b$ and $\theta$ as a contour plot in Fig.~\ref{fig:conshear} (top left). We notice that the critical value of $b$ at $\theta=0$ is $b_c=1$ and this value is independent of magnitude of the magnetic field. In the other extreme, i.e. for $\theta=\pi/2$, the critical value is 
$b_c=[1+B^2/(\varepsilon+P)]^{-1}$
%$1/b_c=1+B^2/(\varepsilon+p)$
, i.e., $b_c$ decreases with increasing magnetic field. In the limit of vanishing magnetic field it has been found in ref.~\cite{Pu:2009fj} that for causal propagation of  the shear modes $b\geq 1$ should be satisfied. In the presence of magnetic field we found that, this constraint can be relaxed to even smaller values of $b$, given that the waves move obliquely. \par
The causality constraint of the fast and slow waves in eq.~\eqref{eq:v_LforSh} can be written in the form of \eqref{Eq:CausBulk}. The simplified expression for the magneto-sonic modes can be written as
\begin{eqnarray}\label{Eq:CausShear}
\text{fast:~~}&&(0<y<1) \land (2\sqrt{y}\leq x<y+1),\nonumber\\
%\text{slow:~~}\alpha +\frac{\sin^{2}{\theta}}{b_{1}}\left(1-\alpha \cos^{2}{\theta}\right\leq 1,\\
\text{slow:~~}&&\left[\left(0<y<1\right)\land \left(x\geq 2\sqrt{y}\right)\right]\lor \left[(y\geq 1)\land (x>y+1)\right].
\end{eqnarray}
We show various causal regions as a function of $b$ and $\theta$ as a contour plot in Fig.~\ref{fig:conshear} (top right and bottom). The critical value of $b$, i.e.,  $b_c=2$ (obtained from \eqref{eq:v_LforSh}) is independent of the angle $\theta$ and the magnitude of magnetic field for the fast magneto-sonic mode. In the absence of a magnetic field this value coincides with that obtained for the sound mode in ref.~\cite{Pu:2009fj}. Similarly, the slow magneto-sonic mode yields the critical value of $b_c=1$, independent of the angle $\theta$ and the magnitude of magnetic field $B$. It is still interesting to see that although the critical values of the fast and slow modes are $B$ and $\theta$ independent, the asymptotic velocities are nevertheless dependent. Increasing the magnetic field, increases the asymptotic group velocities but the causal region always remains causal no matter how large the magnetic field becomes.  

%%%%%%%%%%%%%%%% SHEAR+BULK %%%%%%%%%%%%%%%%%%%%%%%%%%%%%%%%%%%%
\subsection{MHD with both bulk and shear viscosity }
In this subsection, we investigate the stability and causality of a viscous fluid with finite shear and the bulk viscosity  in a magnetic field. 

In heavy-ion collisions the initial magnetic field is very large and both shear and bulk viscosities 
are non-zero for the temperature range achieved in these collisions, hence the present case is 
most relevant to the actual heavy-ion collisions at top RHIC and LHC energies. The energy-momentum tensor is
\begin{equation}\label{eq:EMTensBS}
T^{\mu \nu}=\left(\varepsilon+P+\Pi+B^2\right)u^\mu u^\nu-\left(P+\Pi+\frac{B^2}{2}\right)g^{\mu\nu}-B^{2}b^{\mu}b^{\nu}+\pi^{\mu\nu}.
\end{equation}
The small variation of the energy-momentum tensor due to the perturbed fields is
\begin{eqnarray}\label{eq:perTB+S}
\delta{\tilde{T}}^{\mu\nu}&=\left(\varepsilon_{0}+P_{0}+B_{0}^{2}\right)\left(u_{0}^{\mu}\delta{\tilde{u}}^{\nu}+\delta{\tilde{u}}^{\mu}u_{0}^{\nu}\right)+\left(\delta{\tilde{\varepsilon}}+\delta{\tilde{P}}+\delta{\tilde{\Pi}}+2B_{0}\delta{\tilde{B}}\right)u_{0}^{\mu} u_{0}^{\nu}\notag\\
&-\left(\delta{\tilde{P}}+\delta{\tilde{\Pi}}+B_{0}\delta{\tilde{B}}\right)g^{\mu\nu}-B_{0}^{2}\left(b_{0}^{\mu}\delta{\tilde{b}}^{\nu}+\delta{\tilde{b}}^{\mu}b_{0}^{\nu}\right)-2B_{0}\delta{\tilde{B}}b_{0}^{\mu}b_{0}^{\nu}+\delta \tilde{\pi}^{\mu\nu}.
\end{eqnarray}
Following the same procedure, as discussed in the previous two sections, we obtain the dispersion relations for the following independent variables
\begin{equation}
\delta \tilde{X}=(\delta\tilde{\varepsilon}, \delta\tilde{u}^{x} ,\delta\tilde{u}^{y}, \delta\tilde{u}^{z}, \delta\tilde{b}^{x}, \delta\tilde{b}^{y}, \delta\tilde{\pi}^{xx}, \delta\tilde{\pi}^{xy}, \delta\tilde{\pi}^{xz}, \delta\tilde{\pi}^{yy},\delta\tilde{\pi}^{yz},\delta\tilde{\Pi})^T.
\end{equation}
Following the usual procedure of linearisation we get a $12 \times 12$ dimensional square matrix $A$.
By setting $\det{A}=0$ we have the following equations which subsequently give the dispersion relations
\begin{eqnarray}
\label{eq:non-hBS}
\left(1 + i \omega \tau_{\pi}\right)^2=0,\\
\label{eq:AlfBS}
\omega^3 -\frac{i}{\tau_{\pi}}\omega^2 -\left(v_{A}^2\cos^2{\theta}+\frac{\eta}{h\tau_{\pi}}\right)k^2 \omega + \frac{i}{\tau_{\pi}} k^2 v_{A}^2\cos^2{\theta}=0,\\
\label{eq:MSBS}
\omega^7 + \mathsf{X}_{6}\omega^6 + \mathsf{X}_{5}\omega^5 + \mathsf{X}_{4} \omega^4 +\mathsf{X}_{3}\omega^3 +\mathsf{X}_{2}\omega^2 +\mathsf{X}_{1}\omega +\mathsf{X}_{0}=0,
\end{eqnarray}
where
\begin{eqnarray}
\mathsf{X_6}&=&-i\left(\frac{1}{\tau_{\Pi}}+\frac{2}{\tau_{\pi}}\right),\nonumber
\\
\mathsf{X}_{5}&=&-\frac{1}{\tau_{\pi}}\left(\frac{2}{\tau_{\Pi}}+\frac{1}{\tau_{\pi}}\right)-\left[ v_{A}^2 +\left(\alpha +\frac{1}{b_1}\right)\left(1-v_A^2\sin^2{\theta}\right)+ \frac{1}{3b}\left\{\ 7-v_A^2\left(3+\sin^2{\theta}\right)\right\}\right]k^2,\nonumber\\
%%%%%%%%%%
\mathsf{X}_4 &=& \frac{i}{\tau_{\pi}^2\tau_{\Pi}} + i\biggl[\left(\frac{1}{\tau_{\Pi}}+\frac{2}{\tau_{\pi}}\right) Y + \frac{2}{b_1 \tau_{\pi}}\left(1-v_{A}^2\sin^2{\theta}\right) + \frac{1}{3b}\left(\frac{1}{\tau_{\pi}}+\frac{1}{\tau_{\Pi}}\right)\Big\{7-v_A^2\left(3+\sin^2{\theta}\right)\Big\}\biggr]k^2,\nonumber
\\
\mathsf{X}_3 &=& \biggl[\frac{1}{b_1 \tau_{\pi}^2}\left(1-v_{A}^2\sin^2{\theta}\right)+ \frac{1}{3b\tau_{\pi}\tau_{\Pi}} \left\{\ 7- v_A^2\left(3+\sin^2{\theta}\right) \right\}\  +\left(\frac{2}{\tau_{\Pi}\tau_{\pi}}+\frac{1}{\tau_{\pi}^2}\right) Y   \biggr]k^2 \nonumber\\
& &+ \left[\left(\alpha +\frac{1}{b_1}\right)\left(v_A^2\cos^2{\theta}+\frac{\eta}{h\tau_{\pi}}\right)+\frac{1}{3b}\left\{(v_A^2\left(3+\cos^2{\theta}\right)+\frac{\eta}{3h\tau_{\pi}}\right\} \right]k^4,\nonumber
\\
%%%%%%%%%%%%%%
\mathsf{X}_2&=&-\frac{i}{\tau_{\pi}^2\tau_{\Pi}}\mathsf{Y} k^2-i\Biggl[\frac{1}{b_1\tau_{\pi}}\left(2v_A^2\cos^2{\theta}+\frac{\eta}{h\tau_{\pi}}\right)+\frac{1}{3b\tau_{\Pi}}\left\{ v_A^2 \left(\frac{1}{\tau_{\Pi}}+\frac{1}{\tau_{\pi}}\right)\left(3+\cos^2{\theta}\right)+\frac{4\eta}{3h\tau_{\pi}}\right\}
\nonumber\\
&&+\alpha\left\{\ \frac{\left(1-v_A^2\right)}{b}\left(\frac{1}{\tau_{\Pi}}+\frac{1}{\tau_{\pi}}\right)+v_A^2\left(\frac{1}{\tau_{\Pi}}+\frac{2}{\tau_{\pi}}\right) \right\} \Biggr]k^4,\nonumber
\\
\mathsf{X}_1&=&-\left[\frac{v_{A}^2}{b_1\tau_{\pi}^2}\cos^2{\theta}+\frac{ v_{A}^2}{3b\tau_{\pi}\tau_{\Pi}}\left(3+\cos^2{\theta}\right)+\frac{\alpha\eta}{h\tau_{\pi}\tau_{\Pi}}+\alpha v_A^2 \left(\frac{2}{\tau_{\pi}\tau_{\Pi}}+\frac{1}{\tau_{\pi}^2}\right) \cos^2{\theta} \right] k^2,\nonumber
\\
\mathsf{X}_0&=&\frac{\alpha v_{A}^2}{\tau_{\pi}^2\tau_{\Pi}}k^4 \cos^2{\theta},\hspace*{1.5cm}\mathsf{Y}=  v_{A}^{2} + \alpha \left(1-v_{A}^{2}\sin^{2}{\theta} \right).
\end{eqnarray}
%%%%%%%%
First, we find that eq.~\eqref{eq:non-hBS} gives two non-propagating modes of frequency $\omega=\frac{i}{\tau_{\pi}}$. Now, the eq.~\eqref{eq:AlfBS} is a third-order polynomial and can be solved analytically as 
discussed previously whereas the eq.~\eqref{eq:MSBS} is a seventh-order polynomial equation and can not 
be solved analytically, therefore we lookout for the solution of these equations for some special cases 
discussed below.\par
For $\theta=0$, we obtain two cubic and a single quartic equations. The $X_i$'s of the 
two cubic polynomials are same as in eq.~\eqref{eq:X_iMSSAlf}.
The dispersion relations for these cases are already discussed in the previous section, hence we will not 
repeat them here. The $X_i$'s for the fourth-order polynomial equation are
\begin{align}
X_{3}&=-i\left(\frac{1}{\tau_{\pi}} + \frac{1}{\tau_{\Pi}}\right), &X_{2}&=-\frac{1}{\tau_{\pi}\tau_{\Pi}}-\left(\alpha + \frac{1}{b_{1}} + \frac{4}{3b}\right)k^2,\nonumber\\
X_{1}&=i\left[\alpha \left(\frac{1}{\tau_{\pi}} + \frac{1}{\tau_{\Pi}}\right) + \frac{1}{b_{1}\tau_{\pi}} + \frac{4}{3b\tau_{\Pi}} \right]k^2, & X_{0}&=\frac{\alpha}{\tau_{\pi}\tau_{\Pi}}k^2,
\end{align}
and the corresponding roots can be calculated using the formula given in Appendix~\ref{sec:ployeqns}.\par
For another case, we choose $\theta=\frac{\pi}{2}$, this time two of the roots  turned out to be zero, and another two roots are the same as eq.~\eqref{eq:Pi/2shear}. As before, we call these four modes as shear mode.
The $X_i$'s for the fourth-order polynomial equation are
\begin{align}\label{eq:X-iMSBS}
&X_{0}=\frac{1}{\tau_{\pi}\tau_{\Pi}} v_f^2 k^2,\nonumber\\
& X_{1}=i\left[ \left(\frac{1}{\tau_{\pi}} + \frac{1}{\tau_{\Pi}}\right) v_f^2 + \frac{1}{h\tau_{\pi}\tau_{\Pi}}\left(\zeta +\frac{4}{3}\eta\right)\right]k^2,\nonumber\\
&X_{2}=-\frac{1}{\tau_{\pi}\tau_{\Pi}}-\left[ v_f^2 + \frac{1}{h}\left(\frac{\zeta}{\tau_{\Pi}} + \frac{4\eta}{3\tau_{\pi}}\right)\right]  k^2, \nonumber\\
& X_{3}=-i\left(\frac{1}{\tau_{\pi}} + \frac{1}{\tau_{\Pi}}\right),
\end{align}
and the corresponding roots can be calculated using the formula given in Appendix~\ref{sec:ployeqns}.
\begin{figure}[tbh]
\begin{center}
	\includegraphics[width=0.455\textwidth]{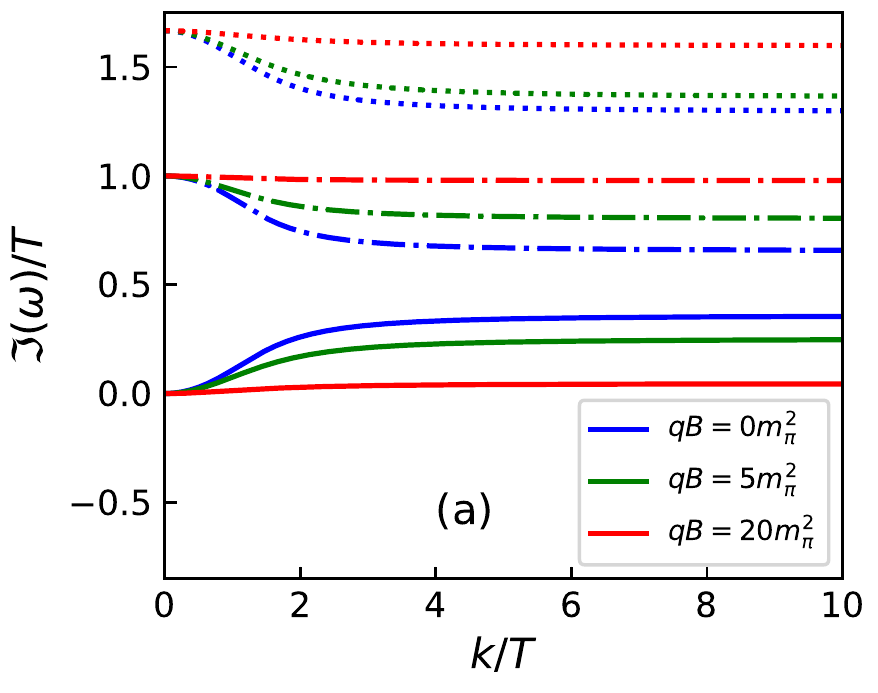}
	\includegraphics[width=0.44\textwidth]{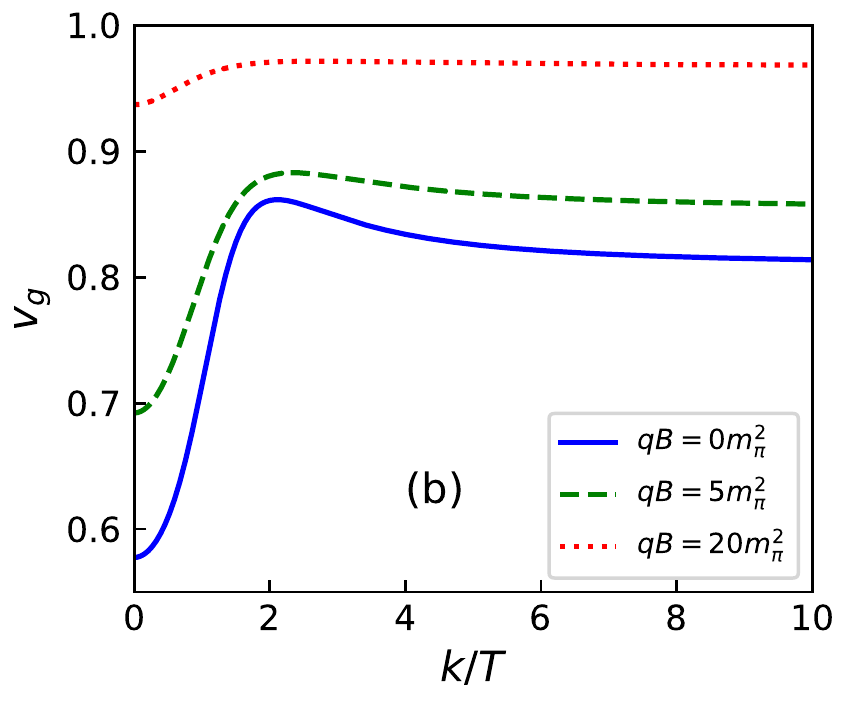}
	\caption{(Color online) In the left panel  $\Im(\omega)/T$ versus $k/T$  and in the right panel group velocity as a function of $k/T$ are plotted for different magnetic fields for $\theta=\frac{\pi}{2}$. $v_g$ is obtained from a quartic equation with the coefficients eq.~\eqref{eq:X-iMSBS}. The solid lines in the left panel corresponds to the propagating modes, the dashed lines and the dash-dotted lines correspond the non-propagating modes. The other parameters used here are \(a=a_{1}=0.1, T=200\, \text{MeV}, \tau_{\Pi}=0.985 \,\text{fm}\ \text{and} \ \tau_{\pi}=0.591\, \text{fm}\) and kept constants for all the curves.}
	\label{fig:StBSImpartplot}
\end{center}	
\end{figure}

Note that the imaginary part of the propagating modes (obtained from eq.~\eqref{eq:X-iMSBS}) are degenerate and hence not shown separately in Fig.~\ref{fig:StBSImpartplot}. The dash-dotted lines in the left panel of Fig.~\ref{fig:StBSImpartplot} correspond to the non-propagating modes generated due to the bulk viscosity, this is because in the small $k$ limit they reduce to $\frac{i}{\tau_{\Pi}}$, and in the same logic the dotted line corresponds to the non-propagating mode due to the shear viscosity. In general, we find that the $\Im(\omega)$ is always positive for our set-up.  So, for this parameter set, the fluid is always stable under small perturbation for non-zero bulk and shear viscosity. Also, we note another interesting point, when the magnetic field is increased the imaginary part of the propagating mode tends to zero i.e, the damping of the perturbation diminishes. \par

\noindent In the small $k$ limit the dispersion relations from eqs.~\eqref{eq:non-hBS}-\eqref{eq:MSBS} become
\begin{eqnarray}
\omega=\left\{
\begin{array}{ll}
\frac{i}{\tau_{\pi}},\\
\frac{i}{\tau_{\Pi}},\\%+{\color{red} \mathcal{O}(k^1),}
%\frac{i}{\tau_{\pi}}-\frac{i\eta}{h}k^2+\mathcal{O}(k^3),\\
\pm k v_{A}\cos{\theta},\\%+\frac{i\eta}{2h}k^2+\mathcal{O}(k^3)
%\frac{i}{\tau_{\pi}}+\mathcal{O}(k^1),\\
\pm k v_M,%+\mathcal{O}(k^2),
\end{array}
\right.
\end{eqnarray}
here also the first root have degeneracy of five. Similarly, in the large $k$ limit using the ansatz $\omega=v_{L} k$, we obtain the  asymptotic group velocities $v_L$ as:
\begin{eqnarray}\label{eq:v_LBS}
v_L^{2}=\left\{\
\begin{array}{ll}
v_A^2\cos^2{\theta}+\frac{\eta}{h\tau_{\pi}},\\
\frac{1}{2}\left[x\pm\sqrt{x^{2} - 4 y}\right],
\end{array}
\right.
\end{eqnarray}
where
\begin{eqnarray}
&&x=\left[v_{A}^{2} + \left(\alpha +\frac{1}{b_{1}} \right) \left(1-v_{A}^{2}\sin^{2}{\theta}\right)  +\frac{1}{3b}\left\{\ 7-v_{A}^{2}\left(3+\sin^{2}{\theta}\right)\right\}\ \right],\nonumber\\
&&y=\left[\left(\alpha +\frac{1}{b_{1}}\right)v_{A}^2 \cos^2{\theta} +\left(\alpha +\frac{1}{b_{1}}+\frac{4}{3b}\right)\frac{\eta}{h\tau_{\pi}}+\frac{v_{A}^2 }{3b}\left(3+\cos^{2}{\theta}\right)\right].
\end{eqnarray}
Now we are ready to explore the causality of a fluid in magnetic field. For this, we again check whether the asymptotic group velocity has super or sub luminal speed. We found that the theory as a whole is causal if the fluid satisfy the following asymptotic causality conditions for magneto-sonic waves:
\begin{eqnarray}\label{Eq:StabRegBandS}
\text{fast:~~}&&(0<y<1) \land (2\sqrt{y}\leq x<y+1),\nonumber\\
\text{slow:~~}&&\left[\left(0<y<1\right)\land \left(x\geq 2\sqrt{y}\right)\right]\lor \left[(y\geq 1)\land (x>y+1)\right].
\end{eqnarray}
\begin{figure}[h!]
\begin{center}
	\includegraphics[width=0.48\textwidth]{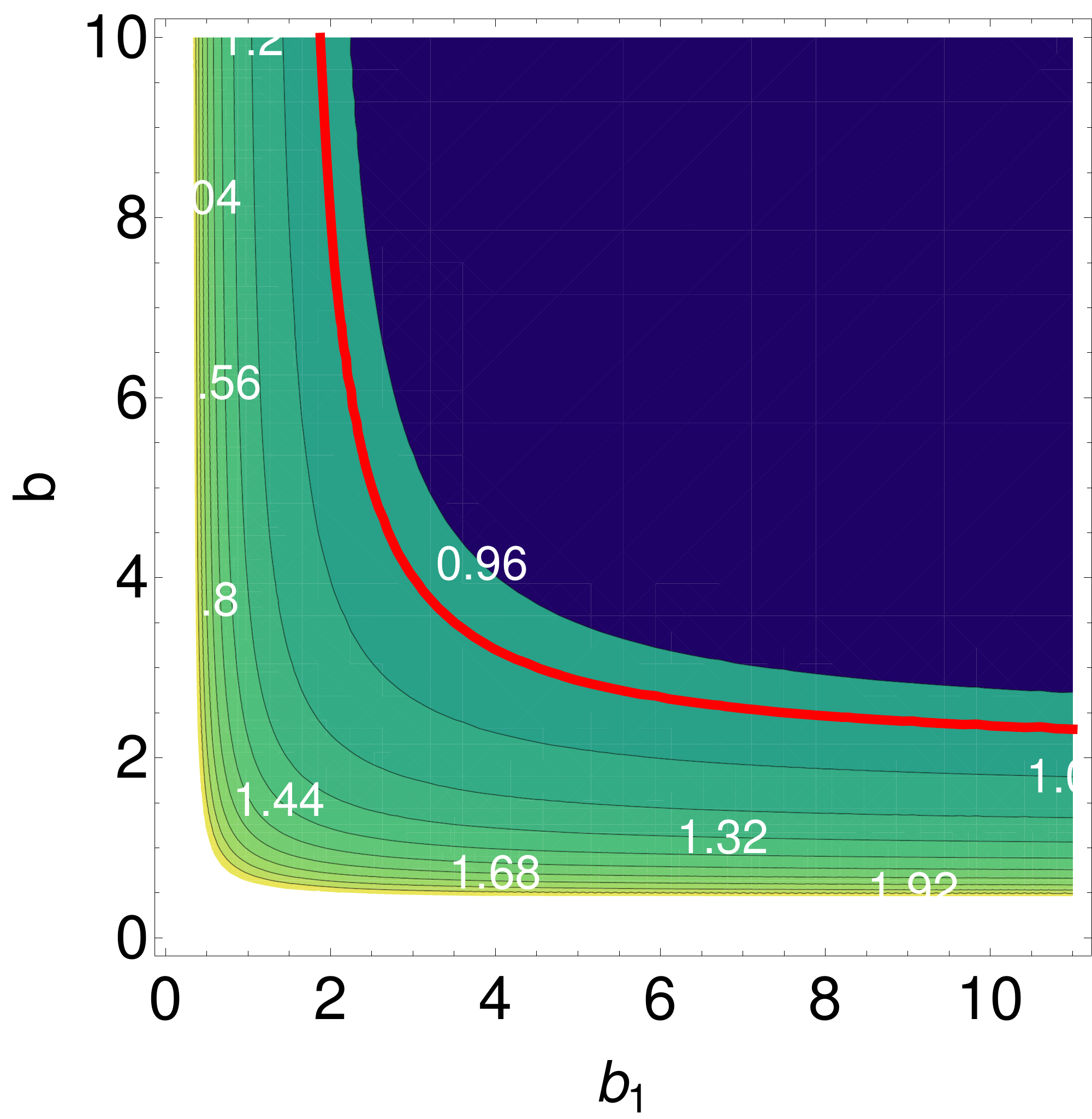}
	\includegraphics[width=0.48\textwidth]{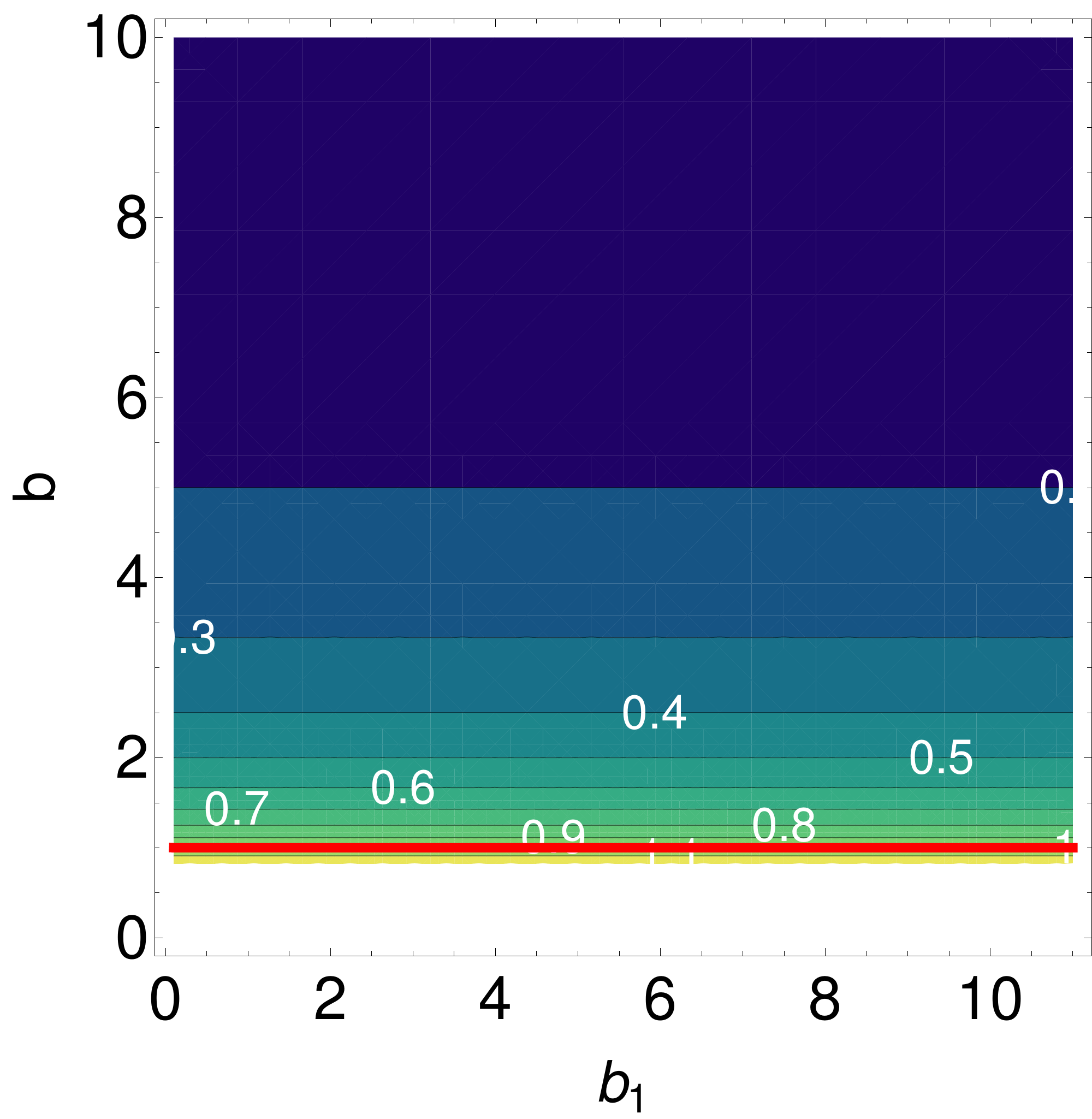}
\caption{(Color online) Contour plot showing various causal regions, obtained from eq.~\eqref{Eq:StabRegBandS}, for fast (left panel) and slow (right panel) branches. The red contour is the critical line of causality, denoting $v_L^2=1$. The region above the red line is causal for the slow magneto-sonic waves and acausal below similarly for the fast magneto-sonic wave right side of red line is causal region and left side is acausal region. The magnitude of the magnetic field has been fixed to $qB=10m^2_{\pi}$ and the other parameters used are $\alpha=1/3$, $T=200\,$MeV.}
	\label{fig:conbS}
\end{center}	
\end{figure}
From eq.~\eqref{eq:v_LBS} we find that a larger magnetic field gives a larger $v_{L}$, but always remain sub-luminal given $b$ and $b_{1}$ are larger than their corresponding critical values (discussed earlier). It is also clear from eq.~\eqref{eq:v_LBS} the asymptotic group velocity for non-zero bulk and shear viscosity is larger than the individual shear and bulk viscous cases.\par
In Fig.~\ref{fig:conbS} we show the contour plot of various causal regions as a function of $b$ and $\theta$. The critical line~(red line) of the fast magneto-sonic mode (left panel) show that  $b$ and $b_1$ are inversely proportional. On the other hand, the causality condition for the slow magneto-sonic waves is independent of $b_1$. The critical value of $b$ for the slow magneto-sonic mode is      $b_c=1$.

%%%%%%%%%%%%%%%%%%%%%%%%%%%%%%%%%%%%%%%%%%%%%
\section{Characteristic velocities for bulk viscosity}
\label{sec5:Charvel}
The characteristic curves can be seen as the lines along which any information is transported in the fluid, for example small perturbations, discontinuities, defects or shocks etc travel along one of these characteristic curves refs.~\cite{Sommerfeld,Hilbert,rezzolla2013relativistic}. Here we take the effect of  non-linearity in the propagation speed which is ignored in the linearisation procedure discussed earlier. Without the loss of generality we consider the~($2+1$)-dimensional case with only bulk viscosity (shear viscosity can be added in the similar way) and write the energy-momentum conservation equations, Maxwell's equations and the IS equation in the standard form for studying the characteristic velocities as
\begin{equation}
P^{\beta}_{mn}\partial_{\beta}Q^{n}+R_{m}=0.
%+X^{x}_{\mu \nu}\partial_{x}+X^{y}_{\mu \nu}\partial_{y}+X^{z}_{\mu \nu}\partial_{z}.
\label{eq:char_mat}
\end{equation}
Here $Q^{n}=\left(\varepsilon, u^{x}, u^{y}, b^{x}, B, \Pi \right)$ and $R_{m}=\left(0,0,0,0, 0, \Pi\right)$. We parametrize the fluid velocity as $u^{\mu}=\left(\cosh{\theta}, \sinh{\theta}\cos{\phi}, \sinh{\theta}\sin{\phi},0\right)$ and the $b^{\mu}=\left(\sinh{\theta}, \cosh{\theta}\cos{\phi}, \cosh{\theta}\sin{\phi},0\right)$. The matrix elements of $P^t_{mn},P^x_{mn},P^y_{mn}$ are given in Appendix~\ref{app:expression}.\par
We find the characteristic velocities ($v_{x}^{ch},v_{y}^{ch}$) by solving the following equations:
\begin{eqnarray}
\det\left(v_{x}^{ch}P^t-P^x\right)=0,\\
\det\left(v_{y}^{ch}P^t-P^y\right)=0.
\label{eq:charac_b}
\end{eqnarray}
For simplicity, here we take fluid in the LRF i.e, $u^{\mu}=\left(1,0,0,0\right)$  and the magnetic filed along the $y$-axis $b^{\mu}=\left(0,0,1,0\right)$. Then the characteristic velocities are
\begin{equation}\label{eq:char_perpen}
v_{x}^{ch}=\pm\sqrt{\frac{B^2+\alpha\left(\varepsilon +P+ \Pi\right)}{\left(h+\Pi\right)}+\frac{\zeta}{\tau_{\Pi}\left(h+\Pi\right)}},
\end{equation}
\begin{equation}
v_{y}^{ch}=\left\{
\begin{array}{ll}
\pm \frac{B}{\sqrt{\left(h+\Pi\right)}},\\\\
\pm\sqrt{\alpha+\frac{\zeta}{\tau_{\Pi}\left(\varepsilon +P+ \Pi\right)}},
\end{array}\label{eq:char_para}
\right.
\end{equation}
where $h=\varepsilon+P+B^2$ and the other roots are zero. The characteristic velocities  obtain in eqs.~\eqref{eq:char_perpen},~\eqref{eq:char_para} are same with the eq.~\eqref{eq:vLbuk} for $\theta=\frac{\pi}{2}$ and $\theta=0$, respectively provide  $\Pi=0$. So we conclude that the asymptotic group velocity obtained by linearizing the MHD-IS equations is same as the characteristic velocities.

%%%%%%%%%%%%%%%%%%%%%%%%%%%%%%%%%%%%%%%%%%%%%

\section{Results from the modified IS theory}
\label{sec6:NRMHD}
In fact, the formulation of second order hydrodynamical theory in the presence of
electromagnetic field is still an active area of research. So far all the results we discussed were obtained for viscous fluid in a magnetic field within the frame-work of the IS theory. The IS relaxation eq.~\eqref{eq:ISshortShear} do not consider the effect of magnetic field. However, recently the authors of ref.~\cite{Denicol:2018rbw} solved the Boltzmann equation in the presence of magnetic field using the 14 moment approximation and found that IS relaxation eq.~\eqref{eq:ISshortShear} gets modified. We call these equations the modified IS equations or the NRMHD-IS (non-resistive MHD-IS) equations. The NRMHD-IS equations shows that the relaxation equation for the shear-stress tensor contains additional terms, here we neglected most of the  terms and only keep the term which couples magnetic field and the shear viscosity. The simplified NRMHD-IS equation takes the following form
\begin{equation}
\tau_{\pi}\frac{d}{d\tau}\pi^{<\mu \nu>}+\pi^{\mu\nu}=2\eta\sigma^{\mu\nu}-\delta_{\pi B} Bb^{\alpha\beta}\Delta^{\mu\nu}_{\alpha\kappa}g_{\lambda\beta}\pi^{\kappa\lambda}.
\label{eq:shearModified}
\end{equation}
Where $\delta_{\pi B}$ is a new coefficient appearing only due to the magnetic field and $b^{\alpha\beta}=-\epsilon^{\alpha\beta\gamma\delta}u_{\gamma}b_{\delta}$ is an anti-symmetric tensor which satisfy $b^{\mu\nu}u_{\nu}=b^{\mu\nu}b_{\nu}=0$. The rank-four traceless and symmetric projection operator is defined as $\Delta^{\mu\nu}_{\alpha\kappa}=\frac{1}{2}\left(\Delta^{\mu}_{\alpha}\Delta^{\nu}_{\kappa}+\Delta^{\mu}_{\kappa}\Delta^{\nu}_{\alpha}\right)-\frac{1}{3}\Delta^{\mu\nu}\Delta_{\alpha\kappa}$. 

Before proceeding further, a few comments on the NRMHD-IS equations are in order.  
It is well known that in the presence of a magnetic field, the transport coefficients split 
into several components, namely three bulk components and five shear components 
refs.~\cite{Grozdanov:2016tdf,Hernandez:2017mch,Denicol:2018rbw,Dash:2020vxk}. 
The  information of these anisotropic transport coefficients are  hidden 
inside the new coupling terms of the modified IS theory eq.~\eqref{eq:shearModified}.
Note that the first-order terms on the right-hand sides are proportional to the usual  shear-viscosity.
% Additional, first-order terms in this equation are also the new term that  couple magnetic field and shear viscosity, which are formally of the first-order in a small quantity of $\pi^{\mu\nu}$. 
These terms can be combined with the first-order terms on  the left-hand side and, 
after inversion of the respective coefficient matrices, will lead to the various anisotropic 
transport coefficients. On the other hand, when solving the full second-order equations 
of the modified IS theory, one does not need to replace the standard viscosity with the 
anisotropic transport coefficients, since the effect of the magnetic field, is already taken
 into account by the new terms in these equations.  Regarding modified second-order 
 theory with finite bulk viscosity, we would like to mention that, there is still no existing 
 theory that yields three distinct bulk components in Navier-Stokes  limit 
 (for details see ref.~\cite{Denicol:2018rbw}) and it is still an open issue. 

The last term of eq.~\eqref{eq:shearModified} is the only non-trivial term  added to the conventional IS theory  for which we already discussed the results in previous sections. So, here we only consider the last term of eq.~\eqref{eq:shearModified} and calculate the corresponding correction to the old results. 

First, we add a perturbation to the new term which contributes to  $\delta \tilde{\pi}^{\mu\nu}$
\begin{eqnarray}
\delta{\tilde{I}}^{\mu\nu}&=&\delta_{\pi B} B_{0}b^{\alpha\beta}_{0}\Delta^{\mu\nu}_{\alpha\kappa}g_{\lambda\beta}\delta{\tilde{\pi}^{\kappa\lambda}}.
\label{eq:Lastterm}
\end{eqnarray}
While calculating eq.~\eqref{eq:Lastterm} we use the fact that in the local rest frame the unperturbed shear stress tensor vanishes i.e., $\pi^{\mu\nu}_{0}=0$, and as a consequence $\delta{\tilde{B}}, \delta{\tilde{b}^{\mu\nu}}$ terms are absent in eq.~\eqref{eq:Lastterm}. For later use we define the projection of a four-vector $A^{\mu}$ as $A^{<\mu>}=\Delta^{\mu}_{\nu}A^{\nu}$, which is orthogonal to $u^{\mu}$.\par
Using these new definitions we write the eq.~\eqref{eq:Lastterm} in a more simplified form as
\begin{eqnarray}
\delta{\tilde{I}}^{\mu\nu}&=&\delta_{\pi B} \frac{ B_{0}}{2}\left(b^{<\mu>}_{\lambda}\delta{\tilde{\pi}^{<\nu>\lambda}} + b^{<\nu>}_{\lambda}\delta{\tilde{\pi}^{<\mu>\lambda}}\right)-\delta_{\pi B} \frac{ B_{0}}{3}\Delta^{\mu\nu}b_{<\kappa>\lambda}\delta{\tilde{\pi}^{<\kappa>\lambda}}.
\label{eq:Lasttermgen}
\end{eqnarray}
In the LRF, 
%$u^{\mu}=(1, 0, 0, 0)$  hence  $\Delta^{\mu\nu}=\Delta_{\mu\nu}=(0, -1, -1, -1)$ and $\Delta^{\mu}_{\nu}=(0, 1, 1, 1)$. 
the following components  of  the $b^{\mu\nu}$  are found to be non-zero $b^{xy}=1$, $b^{yx}=-1$, $b^{x}_{y}=-1$, $b^{y}_{x}=1$, $b^{xy}=b_{xy}=1$ and $b^{yx}=b_{yx}=-1$, where $b^{\mu}$ taken as $(0, 0, 0, 1)$.
For the $(3+1)$ dimensional case  there are  five independent equations for the shear stress according to the IS theory. For each  five equations there are corresponding components of the $\delta \tilde{I}^{\mu\nu}$ which for our case are $\delta \tilde{I}^{xx}=-\delta_{\pi B} B_{0}\delta{\tilde{\pi}^{xy}}$, $\delta \tilde{I}^{xy}=\frac{1}{2}\delta_{\pi B} B_{0}\left(\delta{\tilde{\pi}^{xx}}-\delta{\tilde{\pi}^{yy}} \right)$, $\delta \tilde{I}^{xz}=-\frac{1}{2}\delta_{\pi B} B_{0}\delta{\tilde{\pi}^{yz}}$, $\delta \tilde{I}^{yy}=\delta_{\pi B} B_{0}\delta{\tilde{\pi}^{yx}}$ and $\delta \tilde{I}^{yz}=\frac{1}{2}\delta_{\pi B} B_{0}\delta{\tilde{\pi}^{xz}}$. We include these new terms to the corresponding IS equations that we previously derived in section~(\ref{sec:shear}). Here also we get a $11\times 11$ matrix. As usual, we derive the dispersion relations from $\det({A})=0$ which is a eleventh-order polynomial equation. Since finding the analytic solution of this polynomial equation is not possible, here we investigate some special cases.\par
In the hydrodynamical-limit i.e, in the small $k$ limit we get the following modes
\begin{eqnarray}
\omega=\left\{
\begin{array}{lll}
\frac{i}{\tau_{\pi}},\\%+\mathcal{O}(k^1)
\frac{i}{\tau_{\pi}}\left(1 \pm i B_{0}\delta_{\pi B}\right),\\%+\mathcal{O}(k^1)
\frac{i}{2\tau_{\pi}}\left(2 \pm iB_{0}\delta_{\pi B}\right),\\%+\mathcal{O}(k^1)
\pm v_A k \cos{\theta},\\%+\mathcal{O}(k^2)
\pm v_M k.%+\mathcal{O}(k^2)
\end{array}
\right.
\label{Rissmkalf}
\end{eqnarray}
Note that the frequency of a few non-hydrodynamic modes are changed due to the new coupling terms appearing in the NRMHD-IS theory.\par
For the  large $k$ limit we use the ansatz $\omega=v_L k$ and take only the leading order terms in $k$ which yields the following velocities
\begin{eqnarray}\label{eq:Risv_L}
v_L^{2}=\left\{\
\begin{array}{ll}
v_A^2\cos^2{\theta}+\frac{\eta}{h\tau_{\pi}},\\
\frac{1}{2}\left[x\pm\sqrt{x^{2} - 4 y}\right],
\end{array}
\right.
\end{eqnarray}
here
\begin{eqnarray}
&&x=v_{A}^{2} + \alpha \left(1-v_{A}^{2}\sin^{2}{\theta}\right) +\frac{1}{3b}\left\{\ 7-v_{A}^{2}\left(3\cos^{2}{\theta}+4\sin^{2}{\theta}\right)\right\}\ ,\nonumber\\
&&y=\alpha \left(v_{A}^{2}\cos^{2}{\theta}+\frac{\eta}{h \tau_{\pi}}\right)+\frac{1}{3b}\left\{\ v_{A}^{2}\left(4\cos^{2}{\theta}+3\sin^{2}{\theta}\right)+\frac{4\eta}{h\tau_{\pi}}\right\}\ ,
\end{eqnarray}
and the remaining roots are zero. Since the causality of the fluid depends on the asymptotic causality condition which here is  given in eq.~\eqref{eq:Risv_L} and turned out to be the same as eq.~\eqref{eq:v_LforSh}. So it is clear that the causality condition remains same as eq.~\eqref{Eq:StabRegSH} whereas the dispersion relations get modified.

\section{Conclusions}
\label{sec:conclusion}
The current work goes beyond the previous results of refs.~\cite{Grozdanov:2016tdf,Hernandez:2017mch} which used first order viscous MHD. As is well known the first order gradient terms in the energy-momentum tensor breaks causality, which is reflected from the existence of the superluminal mode.  This prohibits the application of viscous MHD in relativistic systems and it is necessary to have rigorous treatment which the present work aims. The remedy was to go beyond the first viscous corrections in hydrodynamics, and to include second order terms as well. We have studied here the stability and causality of the relativistic dissipative fluid dynamics within the framework of the standard and modified IS theories in the presence of magnetic field. By
linearising the energy-momentum conservation equations, relaxation equations for viscous stresses, and the Maxwell's equations and we have obtained the dispersion relations for various cases. In the absence of viscous stresses, the dispersion relation yields the well-known collective modes namely the Alfv\'en, slow and fast magneto-sonic modes. For the bulk viscous case the Alfv\'en mode turned out to be independent of the bulk viscosity. 
The asymptotic causality constraint for the magneto-sonic modes is independent of the magnetic field and the angle of propagation. For the fast mode, the causality condition was found to be same as that previously derived in ref.~\cite{Denicol:2008ha}
in the absence of magnetic field. The slow mode, on the other hand, remained causal throughout the parameter space. We have also derived the causality bound with finite bulk viscosity using the full non-linear set of the equation using the method of characteristics and found that it agreed with the result obtained using small perturbations. In the presence of shear viscosity, the causality constraint for the
two magnetosonic modes was found to be independent of the magnetic field and the angle of propagation. Shear-Alfv\'en modes, on the other hand, do depend on them. We found that the causality constraint changed in presence of a magnetic field. For the modified IS theory in the presence of shear viscosity, new non-hydrodynamic modes emerged but the causality constraint remained unaltered. Finally, in the presence of both shear and bulk viscosity, we have deduced the causal region of parameter space.    
 
There are many possible directions for future work, namely, the study of causality bounds:~(i) in resistive, second-order dissipative MHD where the electric field is non-zero and contributes in the equations of motion ref.~\cite{Denicol:2019iyh},~(ii) theories which have spin degrees of freedom allows to include effects of polarization and magnetization ref.~\cite{Israel:1978up}. These and other interesting questions will be addressed in  the future. 
\acknowledgments
RB and VR acknowledge financial support from the DST Inspire faculty research grant (IFA-16-PH-167), India. AD, NH, and VR are supported by the DAE, Govt. of India. NH is also supported in part by the SERB-SRG under Grant No. SRG/2019/001680. S.P. is supported by NSFC under Grants No. 12075235. We would also like to thank Ze-yu Zhai for pointing out some typographical errors.

\appendix
\section{Solutions of  dispersion relations}
\label{sec:ployeqns}
In general, the hydrodynamic dispersion relations arise as solutions to
\begin{equation}
 P_n(X_0,X_1,...,X_{n-1})=0,
\end{equation}
where $P = \det A$, is a $n^{\text{th}}$ order polynomial obtained from the determinant of matrix $A$ after linearising the MHD equations. In this appendix, we enlist the roots of certain polynomials $P_n$ that we will encounter throughout this work.
For $n=3$, the polynomial $P_3$ is of the form
\begin{equation}
\omega ^3+ X_2 \omega ^2+X_1 \omega +X_0=0,
\label{eq:ploy3rd}
\end{equation}
and the corresponding roots are given as
\begin{equation}
 \omega_k(X_0,X_1,X_2)=\frac{1}{3} \left(  
 -\frac{\xi^{-(k-1)}\Delta_0}{C} - \xi^{(k-1)} C -X_2
 \right).
\label{eq:3rdployroot1}
\end{equation}
Here $k=1,2,3$, $\xi$ is the primitive cubic root of unity, i.e., $\xi=\frac{-1+\sqrt{-3}}{2}$  and the other variables are defined     
\begin{eqnarray}\label{Eq:C}
 C&=&\sqrt[3]{\frac{{\Delta _1+\sqrt{\left(\Delta_1^2-4 \Delta_0^3\right)}}}{2}},\nonumber\\
\Delta_0&=&X_2^2-3 X_1,\nonumber\\
\Delta_1&=&2 X_2^3-9 X_1 X_2+27 X_0.
\end{eqnarray}
Similarly, for  $n=4$, the polynomial $P_4$ is of the  form
\begin{eqnarray}\label{eq:4thgen}
\omega ^{4}+ X_{3} \omega^3 + X_{2}\omega^2 +  X_{1}\omega +X_{0}=0,
\end{eqnarray}
and the corresponding roots are given as
\begin{eqnarray}
 \omega_{1,2}\left(X_0,X_1,X_2,X_3\right)&=&\pm\frac{1}{2} \sqrt{\left(-2 p+\frac{q}{S}-4
   {S}^2\right)}-{S}-\frac{X_3}{4},\nonumber\\
\omega_{3,4}\left(X_0,X_1,X_2,X_3\right)&=&\pm\frac{1}{2} \sqrt{\left(-2 p-\frac{q}{S}-4
   {S}^2\right)}+{S}-\frac{X_3}{4},
\end{eqnarray}
where
\begin{eqnarray}
 p&=&\frac{1}{8} \left(8 X_2-3 X_3^2\right),\nonumber\\
 q&=&\frac{1}{8} \left(X_3^3-4 X_2 X_3+8 X_1\right),\nonumber\\
 S&=&\frac{1}{2} \sqrt{\left(\frac{1}{3} \left(\frac{\Delta
   _0}{Q}+Q\right)+\frac{1}{12} \left(3 X_3^2-8
   X_2\right)\right)},\nonumber\\
Q&=&\sqrt[3]{\frac{{\Delta _1+\sqrt{\left(\Delta^2_1-4 \Delta_0^3\right)}}}{{2}}},\nonumber\\
\Delta_0&=& X_2^2+12 X_0-3 X_1 X_3,\nonumber\\
\Delta_1&=&2 X_2^3-72 X_0 X_2-9 X_1 X_3 X_2+27 \left(X_1^2+X_0 X_3^2\right).
\end{eqnarray}

\section{Details of matrix $A$ defined in section~\ref{sec:shear} and the characteristic velocities} 
\label{app:expression}
By linearising the energy-momentum conservation equations, Maxwell's equations and IS equation for shear viscosity, we write these in the matrix form as eq.~\eqref{eq:shearAX}. Here the form of matrix $A$ is
\begin{eqnarray}\label{eq:A_for_Sh}
\left(
\begin{array}{cccccccccccc}
 i \omega  & -i k_{x} h & -i k_{y} h & -i k_{z}\left(\varepsilon_{0}+P_{0}\right) & -i \frac{k_{x}}{k_{z}}\omega B_{0}^2 & -i \frac{k_{y}}{k_{z}}\omega B_{0}^2 & 0 & 0 & 0 & 0
   & 0  \\
 -i \alpha  k_{x} & i \omega h  & 0 & 0 & i k_{z} B_{0}^2 \left(\frac{k_{x}^2+k_{z}^2}{k_{z}^2}\right)  & i\frac{k_{x} k_{y}}{k_{z}}B_{0}^2 & -i k_x & -i k_y & -i k_z & 0 &
   0  \\
  -i \alpha  k_y & 0 & i  \omega  h & 0 & i\frac{ k_{x} k_{y}}{k_{z}}B_{0}^2 & i k_{z} B_{0}^2 \left(\frac{k_{y}^2+k_{z}^2}{k_{z}^2}\right) & 0 & -i k_x & 0 & -i k_y & -i
   k_z  \\
 -i \alpha  k_z & 0 & 0 & i \omega \left(\varepsilon_{0}+P_{0}\right) & 0 & 0 & i k_z & 0 & -i k_x & i k_z & -i k_y  \\
 0 & i B_0 k_z & 0 & 0 & i \omega  B_0 & 0 & 0 & 0 & 0 & 0 & 0  \\
 0 & 0 & i B_0 k_z & 0 & 0 & i \omega  B_0 & 0 & 0 & 0 & 0 & 0  \\
 0 & -\frac{4}{3} i \eta  k_x & \frac{2}{3} i \eta  k_y & \frac{2}{3} i \eta  k_z & 0 & 0 & f& 0 & 0 & 0 & 0  \\
 0 & -i \eta  k_y & -i \eta  k_x & 0 & 0 & 0 & 0 & f & 0 & 0 & 0  \\
 0 & -i \eta  k_z & 0 & -i \eta  k_x & 0 & 0 & 0 & 0 & f & 0 & 0  \\
 0 & \frac{2}{3} i \eta  k_x & -\frac{4}{3} i \eta  k_y & \frac{2}{3} i \eta  k_z & 0 & 0 & 0 & 0 & 0 & f & 0 \\
 0 & 0 & -i \eta  k_z & -i \eta  k_y & 0 & 0 & 0 & 0 & 0 & 0 & f 
\end{array}
\right),\nonumber\\
\end{eqnarray}
where $f=1+i\omega \tau_{\pi}$. Similarly we can write the matrix $A$ for the modified IS theory, also for both the bulk and shear viscosity case.\par
In section~\ref{sec5:Charvel} we derive the characteristic velocities for the MHD with the bulk viscosity only. For simplicity we consider~($2+1$)-dimensional case and write the energy-momentum conservation equations, Maxwell's equations and the IS equation for bulk in the form of eq.~\eqref{eq:char_mat}.
The matrix elements of $P_{mn}^{t}$ are
\begin{align*}
P_{11}^t&=(1 + \alpha)\cosh^2{\theta}-\alpha,            &          P_{12}^t&=2\left(\varepsilon + P + \Pi \right)\sinh{\theta}\cos{\phi},
\\
P_{13}^t&=2\left(\varepsilon + P + \Pi \right) \sinh{\theta}\sin{\phi}           &         P_{15}^t&=B ,
\\
P_{16}^t&=\sinh^2{\theta} ,               &              P_{21}^t&=(1 + \alpha)\sinh{\theta}\cosh{\theta}\cos{\phi},
\\
P_{23}^t&=\frac{\sin(2\phi)}{2\cosh{\theta}}\left[\left(\varepsilon + P + \Pi \right)\sinh^2{\theta}-B^2 \right],                  &                P_{24}^t&=-B^2\sinh{\theta},
\\
P_{26}^t&=\sinh{\theta}\cosh{\theta}\cos{\phi} ,               &                  P_{31}^t&=(1+ \alpha)\sinh{\theta}\cosh{\theta}\sin{\phi},
\\
P_{34}^t&=B^2\sinh{\theta}\cot{\phi} ,                   &                           P_{36}^t&=\sinh{\theta}\cosh{\theta}\sin{\phi},
\\
P_{42}^t&=B\sinh{\theta},                    &                 P_{44}^t&=-B\cosh{\theta} ,   
\\
P_{45}^t&=-\cos{\phi},                        &                P_{52}^t&=-B\sinh{\theta}\cot{\phi} ,
\\
P_{54}^t&=B\cosh{\theta}\cot{\phi} ,               &              P_{55}^t&=-\sin{\phi},
\\
P_{62}^t&=\zeta \tanh{\theta}\cos{\phi} ,             &               P_{63}^t&=\zeta \tanh{\theta}\sin{\phi} ,
\\
P_{66}^t&=\tau_{\Pi} \cosh{\theta},
\end{align*}
\begin{align*}
P_{22}^t&=\frac{1}{2\cosh{\theta}}\bigl[2\left(\varepsilon + P + \Pi \right)\bigl(\cosh^2{\theta} +\sinh^2{\theta}\cos^2{\phi}\bigr)+B^2\big\{\cosh(2\theta)-\cos(2\phi)\big\}\bigr],\\
P_{32}^t&=\frac{\cot{\phi}}{2\cosh{\theta}}\bigl[2\left(\varepsilon + P + \Pi \right)\sinh^2{\theta}\sin^2{\phi}-B^2\bigl\{(\cosh(2\theta)-\cos(2\phi)\bigr\}\bigr],\\
P_{33}^t&=\frac{\cos^2{\phi}}{\cosh{\theta}}\left[\left(\varepsilon + P + \Pi \right) \left(\cosh^2{\theta} + \sinh^2{\theta}\sin^2{\phi}\right)+B^2 \right] ,\\
\end{align*}
The matrix elements of $P_{mn}^{x}$ are
\begin{align*}
P_{11}^x&=(1 + \alpha)\sinh{\theta}\cosh{\theta}\cos{\phi},             &                P_{13}^x&=\frac{\sin(2\phi)}{2\cosh{\theta}}\left[\left(\varepsilon + P + \Pi \right)\sinh^2{\theta}-B^2 \right],
\\
P_{14}^x&=-B^2\sinh{\theta},              &             P_{16}^x&=\sinh{\theta}\cosh{\theta}\cos{\phi}, 
\\   
P_{21}^x&=(1 + \alpha)\sinh^2{\theta}\cos^2{\phi} +\alpha,            &            P_{22}^x&=2\left(h+\Pi\right)\sinh{\theta}\cos{\phi},
\\P_{24}^x&=-2B^2\cosh{\theta}\cos{\phi} ,                  &               P_{25}^x&=-B\cos(2\phi),
\\
P_{26}^x&=1 + \sinh^2{\theta}\cos^2{\phi},      &       P_{31}^x&=(1 + \alpha)\sinh^2{\theta}\sin{\phi}\cos{\phi},
\\
P_{33}^x&=\left(\varepsilon + P + \Pi \right)\sinh{\theta}\cos{\phi},      &       P_{34}^x&=B^2\cosh{\theta}\cos(2\phi)\csc{\phi} ,
\\
P_{35}^x&=-B\sin(2\phi) ,             &              P_{36}^x&=\sinh^2{\theta}\sin{\phi}\cos{\phi} ,
\\
P_{52}^x&=-\frac{B\sin{\phi}}{\cosh{\theta}}\left[1+\sinh{\theta}\csc^2{\phi}\right] ,
                   &             P_{53}^x&=\frac{B}{\cosh{\theta}}\cos{\phi} ,
\\
P_{54}^x&=B\sinh{\theta}\csc{\phi},             &         P_{62}^x&=\zeta,
\\
P_{66}^x&=\tau_{\Pi}\sinh{\theta}\cos{\phi},
\end{align*}
\begin{align*}
P_{12}^x&=\frac{1}{2\cosh{\theta}}\bigl[2\left(\varepsilon + P + \Pi \right)\bigl(\cosh^2{\theta}+\sinh^2{\theta}\cos^2{\phi}\bigr)+B^2\big\{\cosh(2\theta)-\cos(2\phi)\big\}\bigr],\\
P_{32}^x&=\left(\varepsilon + P + \Pi \right)\sinh{\theta}\sin{\phi}-B^2\sinh{\theta}\cos(2\phi )\csc{\phi},
\end{align*}
The matrix elements of $P_{mn}^{y}$ are
\begin{align*}
P_{11}^y&=(1+ \alpha)\sinh{\theta}\cosh{\theta}\sin{\phi},              &      P_{14}^y&=B^2\sinh{\theta}\cot{\phi},
\\
P_{16}^y&=\sinh{\theta}\cosh{\theta}\sin{\phi},                 &            P_{21}^y&=(1 + \alpha)\sinh^2{\theta}\sin{\phi}\cos{\phi},
\\
P_{23}^y&=\left(\varepsilon + P + \Pi \right)\sinh{\theta}\cos{\phi}          &                 P_{24}^y&=B^2\cosh{\theta}\cos(2\phi)\csc{\phi},
\\
P_{25}^y&=-B\sin(2\phi),                  &                      P_{26}^y&=\sinh^2{\theta}\sin{\phi}\cos{\phi},
\\
P_{31}^y&=(1 + \alpha)\sinh^2{\theta}\sin^2{\phi}+\alpha,               &            P_{32}^y&=-2B^2\sinh{\theta}\cos{\phi} ,
\\
 P_{33}^y&=2\left(\varepsilon + P + \Pi \right)\sinh{\theta}\sin{\phi} ,            &       
P_{34}^y&=2B^2\cosh{\theta}\cos{\phi} ,
\\
P_{35}^y&=B\cos(2\phi),              &                  P_{36}^y&=1 + \sinh^2{\theta}\sin^2{\phi} ,
\\
P_{42}^y&=\frac{B\sin{\phi}}{\cosh{\theta}}\left[1+\sinh^2{\theta}\csc^2{\phi}\right],
                   &                  P_{43}^y&=-\frac{B}{\cosh{\theta}}\cos{\phi} ,
\\
P_{44}^y&=-B\sinh{\theta}\csc{\phi} ,                &               P_{63}^y&=\zeta ,           
\\
P_{66}^y&=\tau_{\Pi}\sinh{\theta}\sin{\phi} .
\end{align*}
\begin{align*}
P_{12}^y&=\frac{\cot{\phi}}{2\cosh{\theta}}\bigl[2\left(\varepsilon + P + \Pi \right)\sinh^2{\theta}\sin^2{\phi}-B^2\big\{(\cosh(2\theta)-\cos(2\phi)\big\}\bigr],\nonumber\\
P_{13}^y&=\frac{\cos^2{\phi}}{\cosh{\theta}}\bigl[\left(\varepsilon + P + \Pi \right) \bigl(\cosh^2{\theta} + \sinh^2{\theta}\sin^2{\phi}\bigr)+B^2 \bigr],\nonumber\\
P_{22}^y&=\left(\varepsilon + P + \Pi \right)\sinh{\theta}\sin{\phi}-B^2\sinh{\theta}\cos(2\phi )\csc{\phi},
\end{align*}
and all the other coefficients are zero.


\bibliographystyle{JHEP}
\bibliography{MHDstable}

\end{document}